\documentclass[
  aps,          
  prl,          
  reprint,      % 2-column PRL look
]{revtex4-2}
\usepackage{amsthm}
\usepackage{amsmath,amssymb}
\usepackage{graphicx}% Include figure files
\usepackage{dcolumn}% Align table columns on decimal point
\usepackage{bm}% bold math
\usepackage{xcolor}
\usepackage{siunitx}
\usepackage{tikz}
\usepackage{pgfplots}
\usepackage{bbm}
\usepackage[normalem]{ulem}

\usepackage[english]{babel}   \usepackage{braket}

\babeltags{en=english}

\newtheorem{theorem}{Theorem}
\newtheorem{corollary}[theorem]{Corollary}
\newtheorem{lemma}{Lemma}

\usepackage{hyperref}% add hypertext capabilities
\DeclareMathOperator{\Tr}{Tr}

\begin{document}

    \title{Universal Predictors for Mixing Time more than Liouvillian Gap}
    
    \author{Yi-Neng Zhou}
    \affiliation{Department of Theoretical Physics, University of Geneva, 24 quai Ernest-Ansermet, 1211 Genève 4, Suisse}
    \thanks{zhouyn.physics@gmail.com}
	\date{\today}

	\begin{abstract}

We analyze the mixing time of open quantum systems governed by the Lindblad master equation, showing that it is determined not only by the Liouvillian gap, but also by the trace-norm factor of each decaying Liouvillian eigenmode. By utilizing them as universal predictors of mixing time, we establish general conditions for the fast and rapid mixing, respectively. Specifically, we derive rapid mixing conditions for both the strong and weak dissipation regimes, formulated as sparsity constraints on the Hamiltonian and the local Lindblad operators. Our findings provide a general framework for calculating mixing time and offer a guide for designing dissipation to achieve desired mixing speeds, which has significant implications for efficient experimental state preparation. 
    
	\end{abstract}

 \maketitle

     {\em Introduction.---} The study of open quantum systems is of paramount importance, both theoretically and experimentally \cite{Breuer2002,Gardiner2004}. In real-world experiments, the interaction between a quantum system and its environment is almost inevitable. Therefore, it is crucial to model quantum many-body systems as open systems and understand their dynamics, which are closely related to phenomena like information scrambling \cite{48651,2018NatPh..14..988S, Mi_2021,Landsman_2019,Vermersch_2019,Bhattacharya_2022, PhysRevLett.131.160402,PhysRevResearch.5.033085,Zhou_2024_LSM}, quantum chaos \cite{Shenker_2014, Maldacena:2015waa, 1969JETP...28.1200L, Syzranov_2018, Han_2022, 10.21468/SciPostPhys.20.3.072,PhysRevResearch.4.L022039}, and quantum phase transitions \cite{RevModPhys.59.1, Carusotto_2013, Raftery_2014, Schmitt_2020, fang2024probingcriticalphenomenaopen, Zhou_2024,10.21468/SciPostPhysCore.6.1.023}. Furthermore, the dynamics of open systems can be harnessed for dissipative state preparation \cite{PhysRevLett.77.4728, Plenio_1999, Diehl_2008, Kraus_2008, NP2009, PhysRevLett.107.080503, Diehl_2011, Barreiro_2011, Kastoryano_2011, Bardyn_2013, Lin_2013, Shankar:2013xif, Leghtas_2015, Ding_2024,zhan2025rapidquantumgroundstate,lin2025dissipative, PRXQuantum.6.010361}, where carefully designed dissipation acts as a tool to steer a system toward a specific initial state relevant for quantum computation and quantum simulation \cite{Feynman1982, f409a97a-1b7e-38fc-a741-afe96ecf8ad1, RevModPhys.86.153, 2012NatPh...8..267B, Greiner2002, Nielsen_Chuang_2010, Lanyon_2011, Martinez_2016}.

A key concept in the study of open system dynamics is the mixing time, defined as the minimum time it takes for an arbitrary initial state to evolve to the system's steady state \cite{PhysRevA.77.042312, Temme_2010, PhysRevE.92.042143}. In dissipative state preparation, mixing time is a vital metric because it quantifies the practical time required to prepare a desired state, thus determining whether a protocol is experimentally feasible \cite{PhysRevResearch.6.033147,zhan2025rapidquantumgroundstate}. Also, an estimation of the mixing time typically plays a crucial role in many of the quantum algorithmic applications to yield efficient quantum algorithms \cite{Grover1996, Aharonov2001}. Also, mixing time is related to entropy \cite{Huang1987, Levin2009, Diaconis1993}, entanglement dynamics, and the structure of the Lindblad spectrum \cite{PhysRevLett.125.230604,PhysRevResearch.3.043137} and the dissipative phase transition, where at the transition point, the mixing time diverges \cite{Levin2009, PhysRevE.92.042143, Lesanovsky2013, PhysRevA.90.052109, PhysRevA.98.042118, PhysRevB.101.214302, PhysRevA.102.012219}.

Considerable progress has been made in bounding mixing times for open quantum dynamics. For Lindbladians satisfying quantum detailed balance, one can map the dynamics to a local frustration-free parent Hamiltonian on a doubled Hilbert space and use its spectral gap to estimate mixing times~\cite{Diehl2008,Temme_2013,Kastoryano:2016feb}. Modified logarithmic Sobolev inequalities provide a complementary approach via entropy decay and can yield tighter $\mathrm{polylog}(L)$ bounds in favorable cases~\cite{Zegarlinski1990LogSobolevIF,bobkov2006modified,Kastoryano_2013,capel2021modifiedlogarithmicsobolevinequality,Bardet_2021,Bardet_2023,Gao:2021xaw}. Other methods apply in restricted regimes, for instance when coherent and dissipative contributions can be analyzed separately~\cite{Fang_2025}. Since these approaches rely on specific structural properties of the dynamics, a general framework for determining the mixing time of generic Lindbladians remains desirable.

In this Letter, we map the definition of the mixing time of an open quantum system to the distance between wavefunctions in a doubled Hilbert space, thereby directly relating it to the Liouvillian gap and the trace-norm factors of the decaying Liouvillian eigenmodes. This connection allows us to identify general conditions for fast and rapid mixing. We further apply our framework to the weak- and strong-dissipation regimes, where the resulting rapid mixing criteria can be expressed as sparsity constraints on the Hamiltonian and the local Lindblad operators. Overall, our results provide a general route to estimating mixing time and offer practical principles for engineering open-system dynamics for efficient state preparation.

{\em Mixing Time and the Lindblad Spectrum}---We consider the open system whose time evolution is driven by the Lindblad Master equation \cite{Gorini:1975nb, lindblad1976generators}
	\begin{equation} 
		\frac{\partial{\hat{\rho}}}{\partial t}=-i[\hat{H},\hat{\rho}] +2\gamma\sum_{\alpha} \hat{K}_{\alpha} \hat{\rho} \hat{K}^{\dagger}_{\alpha}-\gamma\sum_{\alpha}\lbrace \hat{K}^{\dagger}_{\alpha} \hat{K}_{\alpha}, \hat{\rho}\rbrace.
	\end{equation}
Here, $\gamma$ denotes the dissipation strength and $\hat K_\alpha$ the
Lindblad operator. We write $\hat\rho(t)=e^{\hat{\mathcal L}t}[\hat\rho_0]$ for the density matrix evolved from the initial state $\hat\rho_0$.
%We will use the shorthand $\hat{\rho}(t) = e^{\hat{\mathcal{L}}t}[\hat{\rho}_0]$.

We recall the definition of the mixing time, which quantifies how long it takes for an arbitrary initial state to evolve close to the steady state $\hat{\sigma}$ of the dynamics. Closeness is measured by the trace distance $
D(\hat{\rho},\hat{\sigma})=\frac{1}{2}\|\hat{\rho}-\hat{\sigma}\|_{1}$,
where the trace norm is $\|\hat{A}\|_{1}\equiv \mathrm{Tr}\!\left[\sqrt{\hat{A}^{\dagger}\hat{A}}\right]$. %For Hermitian $\hat{A}$, $\|\hat{A}\|_{1}=\mathrm{Tr}|\hat{A}|$. 
The mixing time at accuracy $\eta$ is then defined as
\begin{equation}
\label{mixing_time_def}
\tau_{\mathrm{mix}}(\eta)=\min\left\{t\,\big|\, D\!\left(e^{\hat{\mathcal{L}}t}[\hat{\rho}_{0}],\hat{\sigma}\right)\le \eta,\ \forall\,\hat{\rho}_{0}\right\}.
\end{equation}
That is, $\tau_{\mathrm{mix}}(\eta)$ is the minimal time such that, for arbitrary initial state $\hat{\rho}_{0}$, the evolved state $e^{\hat{\mathcal{L}}t}[\hat{\rho}_{0}]$ is within trace distance $\eta$ of the steady state $\hat{\sigma}$. 

%Since density matrices are Hermitian, this distance can be written as $D(\hat{\rho}(t),\hat{\sigma})=\frac{1}{2}\,\mathrm{Tr}\,|\hat{\rho}(t)-\hat{\sigma}|$.

We now introduce a mapping that relates the mixing time directly to spectral properties of the Lindbladian. To make this connection precise, we use vectorization, or equivalently the Choi--Jamio{\l}kowski isomorphism~\cite{TysonOperatorSchmidtDecompositionsFourier2003,VidalMixedStateDynamicsOneDimensional2004}, to map a density matrix to a wave-function in a doubled Hilbert space:
\begin{equation}
\begin{split}
\hat{\rho}=\sum_{m,n}\rho_{mn}\,|m\rangle\langle n|
\to
|\psi_{\rho}^D\rangle
=\sum_{m,n}\rho_{mn}\,|m\rangle_L\otimes|n\rangle_R .
\label{double_mapping_density_def}
\end{split}
\end{equation}
Here $L$ and $R$ denote the left and right copies. Under this mapping, the Lindblad master equation becomes a Schr\"odinger-like equation in doubled space \cite{PhysRevResearch.3.043060},
\begin{equation}
    i\partial_t|\psi_\rho^D(t)\rangle
    =
    \hat H^D|\psi_\rho^D(t)\rangle ,
\end{equation}
with non-Hermitian doubled-space generator $ \hat H^D=\hat H_s-i\hat H_d $, with
\begin{equation}
\begin{split}
    \hat H_s
    &=
    \hat H_L\otimes \hat{\mathcal I}_R
    -
    \hat{\mathcal I}_L\otimes \hat H_R^T ,
    \\
    \hat H_d
    &=
    \gamma\sum_\alpha
    \Big[
    -2\hat K_{\alpha,L}\otimes \hat K_{\alpha,R}^*
    +
    (\hat K_\alpha^\dagger\hat K_\alpha)_L
    \otimes \hat{\mathcal I}_R
    \\
    &\hspace{1.5cm}
    +
    \hat{\mathcal I}_L
    \otimes
    (\hat K_\alpha^\dagger\hat K_\alpha)_R^*
    \Big] .
\end{split}
\label{eq:doubled_hamiltonian_def}
\end{equation}
Here $T$ denotes transpose and $^*$ denotes complex conjugation. Let $|\psi_{\sigma_j}\rangle$ to be the right eigenstate of $\hat H^D$, with eigenvalue $
    \epsilon_j=\alpha_j-i\beta_j$($\beta_j\geq 0$). Equivalently, the corresponding Liouvillian eigenvalue is
$-i\epsilon_j=-\beta_j-i\alpha_j$. The real numbers $\beta_j$ are decay rates, while $\alpha_j$ generate oscillations. The steady state corresponds to $\beta_0=\alpha_0=0$ \footnote{Trace preservation implies $ \Tr \hat\sigma_j=0, j\neq 0 .$ Hermiticity preservation implies that eigenoperators occur in conjugate
pairs: if $\hat\sigma_j$ has eigenvalue $\lambda_j$, then
$\hat\sigma_j^\dagger$ has eigenvalue $\lambda_j^*$.}. The Liouvillian gap is definded to be the smallest nonzero decay rate $\Delta = \min_{j\neq 0}\beta_j$. Previous studies have emphasized the connection between
$\tau_{\rm mix}$ and $\Delta$~\cite{Temme_2010,Kastoryano_2013,Brand_o_2015}. Below, we show that the gap controls only the exponential decay rate; the mixing time also depends on the trace-norm factor of the Liouvillian decaying eigenmodes.

First consider the simplest case in which the initial state has only one decaying eigenmode $\hat{\sigma}_1$: $|\psi_{\rho_0}^D\rangle
    =
    |\psi_{\sigma_0}\rangle
    +
    c_1|\psi_{\sigma_1}\rangle$ with $\beta_1=\Delta$. Here, $\hat{\sigma}_0$ is the steady state, and $\hat{\sigma}_1$ is slowest decaying eigenmode. The time-evolved deviation is
then $|\psi_{\rho_0}^D(t)\rangle-  |\psi_{\sigma_0}\rangle =
    c_1e^{-i\alpha_1 t-\Delta t}|\psi_{\sigma_1}\rangle,$
and the trace distance is
\begin{equation}
    D(\hat\rho(t),\hat\sigma_0)
    =
    \frac{|c_1|}{2}
    e^{-\Delta t}
    \|\hat\sigma_1\|_1 .
    \label{eq:doubled_single_mode_trace_distance}
\end{equation}
The relaxation time defined by
$D(\hat\rho(T_\eta),\hat\sigma_0)=\eta$ is $T_\eta
    =
    \frac{1}{\Delta}
    \log\left(
    \frac{|c_1|\|\hat\sigma_1\|_1}{2\eta}
    \right)$. Thus, the eigenmode trace-norm factor enters only logarithmically. In this single-mode case, physicality already bounds the prefactor. Since
$\hat\rho_0$ and $\hat\sigma_0$ are normalized density, $ D(\hat\rho_0,\hat\sigma_0) =\frac{|c_1|}{2}\|\hat\sigma_1\|_1 \leq 1 $, so that $|c_1|\|\hat\sigma_1\|_1\leq 2$. A physical initial state supported on a single decaying eigenmode therefore cannot produce a
parametrically large prefactor.

The situation changes for a generic initial state,
\begin{equation}
    |\psi_{\rho_0}^D\rangle
    =
    |\psi_{\sigma_0}\rangle
    +
    \sum_{j>0}c_j|\psi_{\sigma_j}\rangle .
\end{equation}
The initial deviation is $|\psi_{X_0}^D\rangle
    =
    |\psi_{\rho_0}^D\rangle
    -
    |\psi_{\sigma_0}\rangle
    =
    \sum_{j>0}c_j|\psi_{\sigma_j}\rangle $,
or equivalently $ \hat X_0
    =
    \hat\rho_0-\hat\sigma_0
    =
    \sum_{j>0}c_j\hat\sigma_j$.
Physicality imposes only the global constraint
\begin{equation}
    \|\hat X_0\|_1
    =
    \|\hat\rho_0-\hat\sigma_0\|_1
    \leq 2 .
    \label{eq:doubled_full_deviation_bound}
\end{equation}
It does not separately bound the individual modal contributions
$\|c_j\hat\sigma_j\|_1$. These components may be large and cancel in the
full physical operator $\hat X_0$. Hence, for multi-mode initial states,
the relevant object is not a single mode expansion coefficient $c_j$, but the
trace-norm size of the projection of $\hat X_0$ onto each decaying eigenmode.

For clarity, we first assume that $\hat H^D$ is diagonalizable. We defined its left eigenvectors to be $(\hat H^D)^\dagger|\psi_{\tau_j}\rangle
    =
    \epsilon_j^*|\psi_{\tau_j}\rangle$ with orthogonal condition to be $\langle\psi_{\tau_i}|\psi_{\sigma_j}\rangle
    =
    \Tr(\hat\tau_i^\dagger\hat\sigma_j)
    =\delta_{ij}$. 
The rank-one doubled-space spectral projector onto the $j$-th decaying eigenmode is $\hat P_j^D=
    |\psi_{\sigma_j}\rangle
    \langle\psi_{\tau_j}|$. Equivalently, as a superoperator acting on ordinary operators, this projector takes the form
\begin{equation}
    \hat P_j(\hat X)
    =
    \hat\sigma_j
    \Tr(\hat\tau_j^\dagger\hat X).
    \label{eq:doubled_single_projector_operator_form}
\end{equation}

We define the trace-norm factor of the mode $\hat{\sigma}_j$ to be
\begin{equation}
    C_j
    =
    \|\hat P_j\|_{1\to 1}
    \equiv
    \sup_{\|\hat X\|_1=1}
    \|\hat P_j(\hat X)\|_1 .
    \label{eq:doubled_single_prefactor_def}
\end{equation}
This is the induced trace norm of the spectral projection as a superoperator;
it should not be confused with the Hilbert-space norm of
$\hat P_j^D$. Using Eq.\eqref{eq:doubled_single_projector_operator_form}, we have
\begin{equation}
    \|\hat P_j(\hat X)\|_1
    =
    \|\hat\sigma_j\|_1
    \left|
    \Tr(\hat\tau_j^\dagger \hat X)
    \right| .
\end{equation}
Using the duality between the trace norm and the operator norm, $\sup_{\|\hat X\|_1=1}
    \left|
    \Tr(\hat\tau_j^\dagger \hat X)
    \right|
    =
    \|\hat\tau_j\|_\infty$, we obtain 
\begin{equation}
    C_j
    =
    \|\hat\sigma_j\|_1
    \|\hat\tau_j\|_\infty .
    \label{eq:doubled_single_prefactor_identity}
\end{equation}
Combining Eq.\eqref{eq:doubled_single_prefactor_def} with constraint in Eq.~\eqref{eq:doubled_full_deviation_bound} gives
\begin{equation}
    \|\hat P_j\hat X_0\|_1
    \leq
    2C_j
\end{equation}
for every physical initial state. The contribution of the decaying eigenmode $\hat{\sigma}_j$ to the evolved trace distance is therefore bounded by
\begin{equation}
    \frac12
    \left\|
    \operatorname{unvec}\!\left(
    e^{-i\hat H^D t}
    \hat P_j^D|\psi_{X_0}^D\rangle
    \right)
    \right\|_1
    \lesssim
    C_j e^{-\beta_j t}.
    \label{eq:doubled_single_mode_contribution_bound}
\end{equation}
Here, $\operatorname{unvec}$ denotes the inverse of the vectorization map in Eq.~\eqref{double_mapping_density_def}, which maps a doubled-space vector back to the corresponding operator. Summing over all decaying eigenmodes gives
\begin{equation}
    D(\hat\rho(t),\hat\sigma_0)
    \lesssim
    \sum_{j:\,\beta_j>0}
    C_j e^{-\beta_j t}.
    \label{eq:doubled_all_decaying_modes_bound}
\end{equation}
Thus, a conservative sufficient upper bound on the
mixing time is
\begin{equation}
    \tau_{\rm mix}(\eta)
    \lesssim
    \max_{j:\,\beta_j>0}
    \frac{1}{\beta_j}
    \log\left(
    \frac{N_{\rm mode} C_j}{\eta}
    \right),
    \label{eq:doubled_mode_mixing_time_bound}
\end{equation}
up to constants of order unity. Here, $N_{\rm mode}$ denotes the number of nonzero Liouvillian eigenmodes \footnote{If Jordan blocks are present, the
right-hand side acquires additional polynomial factors in $t$, and the
resulting upper bound may contain extra logarithmic corrections associated
with the Jordan-block structure.}. Equation~\eqref{eq:doubled_mode_mixing_time_bound} is the central point. The
Liouvillian gap $\Delta$ fixes only the slowest exponential decay rate. The mixing time also depends on the trace-norm factor $C_j$ of the Liouvillian decay eigenmodes. Therefore, a gap-only criterion is incomplete: an eigenmode with decay rate $\beta_j>\Delta$ can still influence the mixing time if its spectral projector carries a sufficiently large factor $C_j$.

A coarse-grained formulation in terms of decay-rate bands \footnote{When several eigenmodes have comparable decay rates, it is often convenient to
group them into decay-rate bands. This replaces the rank-one projectors
$\hat P_j$ by band projectors $\hat P_a=\sum_{j\in\mathcal B_a}\hat P_j$
and the mode prefactors $C_j$ by $C_a=\|\hat P_a\|_{1\to 1}$. The
single-mode discussion in the main text is recovered by taking each band to
contain one eigenmode. See SM for details.} is given in the SM~\cite{SM}. The usual slow-subspace picture is recovered as a special
case \footnote{For a fixed constant $\kappa>1$, define
\begin{equation}
    \mathcal S_{\rm slow}(\kappa)
    =
    \operatorname{span}
    \left\{
    |\psi_{\sigma_j}\rangle:
    \Delta\leq \beta_j\leq \kappa\Delta
    \right\}.
    \label{eq:doubled_slow_subspace_def}
\end{equation}
This subspace controls the late-time dynamics only when all faster modes are
already below the target accuracy. This requires control of both their decay
rates $\beta_j$ and trace-norm prefactors $C_j$.}.
The SM~\cite{SM} also gives a simpler, stronger sufficient condition based on
bounding the trace-norm condition number of the full Liouvillian eigenbasis. 

\emph{Requirements for fast and rapid mixing.---} We first recall two notions of efficient convergence to the steady state. The dynamics is \emph{fast mixing} if the mixing time grows at most polynomially with the system size $L$,
$\tau_{\mathrm{mix}}=\mathcal{O}(\mathrm{poly}(L))$, and \emph{rapid mixing} if it grows at most polylogarithmically,
$\tau_{\mathrm{mix}}=\mathcal{O}(\mathrm{poly}[\log L])$ \cite{Kastoryano_2013, Brand_o_2015, zhan2025rapidquantumgroundstate}.

Our general bound in Eq.\eqref{eq:doubled_mode_mixing_time_bound} immediately translates these notions into requirements on the Liouvillian gap and the trace-norm factor $C_j$ of each decaying eigenmode. We now state general sufficient conditions for fast and rapid mixing. The mixing time is controlled by two ingredients: the inverse Liouvillian gap, which sets the slowest exponential decay scale, and the trace-norm factor of the decaying eigenmodes.

For fast mixing, no assumption on the collective sum over decay eigenmodes is required. A conservative triangle-inequality estimate in Eq.\eqref{eq:doubled_mode_mixing_time_bound} can generate an additional contribution $\Delta^{-1}\log N_{\rm mode}$ to the mixing time. However, the number of decay eigenmodes is bounded by the Liouville-space
dimension $ N_{\rm mode}\leq N^2 -1$, where $N=d_0^L$ is the Hilbert-space dimension ($d_0$ is local Hilbert space dimension). Hence
this contribution remains $\operatorname{poly}(L)$ whenever $\Delta^{-1}=O[\operatorname{poly}(L)]$.

\begin{theorem}[Fast-mixing condition]\label{thm:fast_mixing}
Fast mixing follows if the following conditions both hold:
\begin{enumerate}
    \item The inverse Liouvillian gap grows at most polynomially with system size,
    \begin{equation}
        \Delta^{-1}
        \leq
        \mathcal O\!\left[\operatorname{poly}(L)\right].
        \label{eq:fast_gap_condition}
    \end{equation}

    \item For every decaying eigenmode $\sigma_j$, its trace-norm factor $C_j$ defined in Eq.\eqref{eq:doubled_single_prefactor_def} satisfies
    \begin{equation}
        \log\left(C_j\right)
        \leq
        \mathcal O\!\left[\operatorname{poly}(L)\right].
        \label{eq:fast_prefactor_condition}
    \end{equation}
\end{enumerate}
\end{theorem}

The requirement for rapid mixing is stronger. Since $\log N_{\rm mode}$ can
scale linearly with $L$, the same conservative bound in Eq.\eqref{eq:doubled_mode_mixing_time_bound} may be incompatible
with polylogarithmic mixing. We therefore separate the individual eigenmode factors $C_j$ from the additional collective amplification generated by
summing over decay eigenmodes. Let $A_{\rm coll}(L)$ denote this additional
trace-norm amplification factor. Rapid mixing does not require $A_{\rm coll}(L)$ to be system-size independent; it only requires its logarithmic contribution to the mixing time to remain bounded by a polylogarithmic function of the system size.

\begin{theorem}[Rapid-mixing condition]\label{thm:rapid_mixing}
Let $A_{\rm coll}(L)$ denote the additional trace-norm amplification factor
generated by the collective sum over decay eigenmodes beyond the individual mode factors $C_j$. We assume that its logarithmic contribution to the
mixing time remains polylogarithmic, $\log A_{\rm coll}(L)
    \leq
    \mathcal O\!\left[\operatorname{poly}(\log L)\right]$. Then rapid mixing follows if the following conditions both hold:
\begin{enumerate}
    \item The inverse Liouvillian gap grows at most polylogarithmically,
    \begin{equation}
        \Delta^{-1}
        \leq
        \mathcal O\!\left[\operatorname{poly}(\log L)\right],
        \label{eq:rapid_gap_condition}
    \end{equation}

    \item For every decaying eigenmode $\sigma_j$, its trace-norm factor $C_j$ satisfies
    \begin{equation}
    \log\left(C_j\right)
        \leq
        \mathcal O\!\left[\operatorname{poly}(\log L)\right].
        \label{eq:rapid_prefactor_condition}
    \end{equation}
\end{enumerate}
\end{theorem}

For rapid mixing, the assumption $\log A_{\rm coll}(L)
    \leq
    \mathcal O\!\left[\operatorname{poly}(\log L)\right]$ is natural and is violated only in highly atypical cases. Violating this condition would require contributions from many decay modes to add coherently in trace norm. Such collective amplification depends on the simultaneous alignment of initial-state overlaps, trace norm factor of eigenmodes, phases, and singular-vector structures of many non-normal modes. In SM~\cite{SM}, we argue that this behavior is not expected for generic physical initial states. Accordingly, our rapid-mixing criterion explicitly assumes that the sum over decay sectors does not produce an additional trace-norm amplification beyond the individual factors $C_j$ larger than $\exp\{\mathcal O[\operatorname{poly}(\log L)]\}$.

The resulting sufficient requirements for fast and rapid mixing are summarized in Table~\ref{tab:mixing-regimes}.
\begin{table}[t]
\centering
\begin{tabular}{|c|c|c|}
\hline
$\tau_{\mathrm{mix}}$ & $\Delta^{-1}$ & $\log(C_j)$ \\
\hline
fast mixing  & $ \le\mathrm{poly}(L)$ & $\le\mathrm{poly}(L)$ \\
\hline
rapid mixing & $ \le\mathrm{poly}[\log(L)]$ & $ \le\mathrm{poly}[\log(L)]$ \\
\hline
\end{tabular}
\caption{Conditions for fast and rapid mixing in terms of the Liouvillian gap $\Delta$ and the trace-norm factor $C_j$ of each decaying Liouvillian eigenmode $\hat{\sigma}_j$, as defined in Eq.~\eqref{eq:doubled_single_prefactor_def}. }
\label{tab:mixing-regimes}
\end{table}
We next apply these \emph{general} criteria to the weak- and strong-dissipation regimes \footnote{By strong or weak dissipation, we mean the dimensionless ratio between the typical dissipative rate $\gamma$ and the relevant coherent energy scale $J_{\rm coh}$ of the Hamiltonian. The strong-dissipation regime is $\gamma/J_{\rm coh}\gg 1$, where the dissipator gives the dominant zeroth-order dynamics and the Hamiltonian is treated perturbatively. The weak-dissipation regime is $\gamma/J_{\rm coh}\ll 1$, where the Hamiltonian dominates and the dissipator is the perturbation. Here $J_{\rm coh}$ may denote, depending on the model, a hopping amplitude, interaction strength, or many-body bandwidth.}, where they simplify to explicit, checkable \emph{sufficient} conditions expressed in terms of the Hamiltonian and Lindblad operators. The resulting rapid mixing conditions are summarized in Table~\ref{tab:dissipation-regimes}. We below assume that any additional trace-norm amplification from the collective sum over decay modes contributes at most polylogarithmically to the mixing time.

\begin{table*}[t]
\centering
\begin{tabular}{c c c}
\hline\hline
dissipation & Hamiltonian $\hat{H}$ & Lindblad operator $\hat{K}$ \\
\hline
weak &
$\hat{H}$ gapped &
$\exists\,\alpha \ge \ln{d_0} \ \text{s.t.}\ 
\forall\,|E_s\rangle,\ 
N\!\bigl(\alpha,\hat{K},|E_s\rangle\bigr) \le \mathrm{poly}(L)$ \\
strong &
$\exists\,\alpha \ge \ln{d_0}\ \text{s.t.}\ 
\forall\,|\epsilon_s\rangle,\ 
N\!\bigl(\alpha,\hat{H},|\epsilon_s\rangle\bigr) \le \mathrm{poly}(L)$ &
boundary dissipation \\
\hline\hline
\end{tabular}
\caption{Sufficient conditions for rapid mixing in the weak- and strong-dissipation regimes, expressed as sparsity constraints on the Hamiltonian $\hat{H}$ and the local Lindblad operator $\hat{K}$. For a fixed eigenstate $|\epsilon_s\rangle$ of $\{\hat{K}_j\}$, $N(\alpha,\hat{H},|\epsilon_s\rangle)$ counts the number of indices $p$ for which $|H_{sp}|\ge e^{-\alpha L}$. Likewise, for a fixed eigenstate $|E_s\rangle$ of $\hat{H}$, $N(\alpha,\hat{K},|E_s\rangle)$ counts the number of indices $p$ for which $|K_{sp}|\ge e^{-\alpha L}$. Here, $d_0$ denotes the local Hilbert-space dimension. We further assume that any additional trace-norm amplification from the collective sum over decay modes contributes at most polylogarithmically to the mixing time.
}
\label{tab:dissipation-regimes}
\end{table*}

\emph{Strong-dissipation regime.---} In this regime, the coherent Hamiltonian term is perturbative. The trace-norm
prefactor of each eigenmode is therefore bounded by the square of the trace-norm of right eigenmode, with only small corrections due to the weak non-orthogonality between left and right eigenmodes generated by the Hamiltonian perturbation \footnote{In the absence of the Hamiltonian, and for Hermitian jump operators, the purely dissipative Liouvillian can be written as
\begin{equation}
    \mathcal L_0[\hat X]
    =
    -\frac{1}{2}\sum_\mu \gamma_\mu
    [\hat L_\mu,[\hat L_\mu,\hat X]],
\end{equation}
which is self-adjoint under the Hilbert--Schmidt inner product. Hence its left
and right eigenoperators can be chosen identical and orthonormal,
$\Tr(\hat\sigma_i^\dagger\hat\sigma_j)=\delta_{ij}$. The corresponding
rank-one spectral projector satisfies
\begin{equation}
    \|\mathcal P_j\|_{1\to1}
    \leq
    \|\hat\sigma_j\|_1\|\hat\sigma_j\|_\infty
    \leq
    \|\hat\sigma_j\|_1^2 .
\end{equation}
A weak Hamiltonian perturbation makes the Liouvillian weakly non-normal, so the left and right eigenoperators become slightly non-orthogonal. This produces only perturbative corrections to the above trace-norm prefactor, as long as the perturbation remains controlled in the relevant spectral subspace.}. For Hilbert--Schmidt-normalized eigenoperators $ \|\hat\sigma_i\|_1
    \leq
    \sqrt{N}\,\|\hat\sigma_i\|_2=
    \sqrt{N},$ with $N=d_0^L$.
Therefore, the trace-norm factor of each eigenmode in Eq.~\eqref{eq:doubled_single_prefactor_def} grows at most exponentially in
$L$, so that $\log\left(C_a/ \eta \right)
    \leq
    \mathcal O(L)+\log\left(1/\eta\right).$
For fixed accuracy $\eta$, the fast-mixing condition Eq.~\eqref{eq:fast_prefactor_condition} for $C_j$ is automatically satisfied. Thus, in the strong-dissipation regime, fast mixing is controlled by the scaling of the Liouvillian gap. We estimate the Liouvillian gap using degenerate perturbation theory, leading to the following sufficient condition.

\begin{corollary}
\label{thm:strong_boundary_corollary}
Consider a one-dimensional open quantum system governed by a Lindblad equation with Hermitian jump operators. In the strong boundary-dissipation regime, the dynamics is fast mixing provided that the boundary Lindblad operators do not commute with the system Hamiltonian.
\end{corollary}

For boundary dissipation, the unperturbed Liouvillian has a highly degenerate steady-state manifold: because the dissipator acts only at the edges, changes in bulk degrees of freedom do not shift the zeroth-order eigenvalue. A Hamiltonian perturbation $\hat{V}^D$ therefore has two distinct effects: it (i) mixes different zeroth-order steady states via bulk terms, and (ii) couples the steady manifold to decaying modes through boundary terms. We assign typical scales to the relevant matrix elements, $
    \langle \psi_{0k}^L|\hat V^D|\psi_{1\alpha}^R\rangle \sim J, \ 
    \langle \psi_{0k}^L|\hat V^D|\psi_{0l}^R\rangle \sim J'$. The states $\{|\psi_{0k}^{L,R}\rangle\}$ span the degenerate zeroth-order
steady-state manifold, chosen to diagonalize $\hat V^D$ within that subspace, and $\{|\psi_{1\alpha}^{R}\rangle\}$ are zeroth-order decaying modes. Here, $J$ sets the coupling from the steady-state manifold to decaying modes, while $J'$ sets the splitting scale within the degenerate manifold.

The leading nonzero decay rate of the perturbed steady manifold arises at second order via virtual transitions into the excited sector. Standard degenerate perturbation theory gives the effective eigenvalue
\begin{equation}
E_{0k}\approx E^{(0)}_{0k}+V^{D}_{0k,0k}
+\sum_{q}\frac{|V^{D}_{0k,1q}|^{2}}{E^{(0)}_{0k}-E^{(0)}_{1q}} .
\label{GS_energy_correction}
\end{equation}
Since the smallest diagonal shift $V^D_{0k,0k}$ is zero \footnote{
In the vectorized representation, one has
$\hat{V}^{D}=\hat{H}\otimes\hat{\mathbb{I}}-\hat{\mathbb{I}}\otimes \hat{H}^{T}$, which acts as the commutator superoperator,
$\hat{V}^{D}\,|\hat{\rho}\rangle\rangle = |[H,\hat{\rho}]\rangle\rangle$.
Therefore, any operator $\hat{\rho}$ that commutes with $\hat{H}$ is a zero mode of $\hat{V}^{D}$. In particular, the identity operator (and more generally any function of $\hat{H}$, or any $|m\rangle\langle m|$ in the energy basis) satisfies $[H,\hat{\rho}]=0$, implying an exact eigenvalue $\lambda=0$ of $\hat{V}^{D}$. Since our $\{|\psi_{0k}^{R}\rangle,\langle\psi_{0k}^{L}|\}$ are chosen to diagonalize $\hat{V}^{D}$ within the zeroth-order degenerate manifold, the diagonal matrix elements $\langle\psi_{0k}^{L}|\hat{V}^{D}|\psi_{0k}^{R}\rangle$ coincide with these eigenvalues; hence the smallest value by magnitude is necessarily $\min_k|\lambda_k|=0$.}, thus the Liouvillian gap is set by the dissipative contribution from the second-order term,
\begin{equation}
\Delta  \sim \sum_{q}\frac{|V^{D}_{0k,1q}|^{2}}{E^{(0)}_{1q}-E^{(0)}_{0k}}
\sim c\,\frac{J^{2}}{\gamma_{\mathrm{loc}}},
\end{equation}
with an $\mathcal{O}(1)$ constant $c$ (for our boundary coupling structure, $c=8$)\footnote{The factor $8$ counts: (i) action on the left or right copy, (ii) action at the left or right boundary, and (iii) raising or lowering the boundary state, giving $2^3=8$ possibilities.}. Here, $\gamma_{\mathrm{loc}}$ denotes the local dissipation strength, which we assume to be spatially uniform for simplicity. Crucially, $\Delta$ does not shrink with $L$: only Hamiltonian terms involving boundary sites connect the steady manifold to decaying modes, so the number of contributing excited states is set by the boundary size, i.e., $\mathcal{O}(1)$ in 1D.

Moreover, the first-order correction to the lowest excited mode generically spreads over the entire degenerate zeroth-order manifold. Since that manifold contains $d_0^{2L-2}$ basis states $\log\!\bigl(\mathrm{Tr}|\hat{\sigma}_{1}|\bigr)\le
\log\!\bigl(\sqrt{d_0^{2L-2}}\bigr)\sim \mathcal O(L)$, and this bound is saturated here because the $\hat{V}^D$-diagonal basis typically forms superpositions across the full manifold.
Therefore,
$\tau_{\mathrm{mix}}(\eta)\sim \frac{\gamma_{\mathrm{loc}}}{c\,J^{2}}\left[\log (d_0) \right] \,L$, which implies fast mixing. In  SM~\cite{SM}, we analyze the strong bulk-dissipation regime and show that this perturbative mechanism does not apply. However, this does not rule out the possibility of rapid mixing in other strong bulk-dissipation settings. Physically, boundary-driven fast mixing is a quantum Zeno stabilization mechanism: strong edge dissipation sets an $\mathcal O(1)$ scale, while boundary Hamiltonian couplings generate a size-independent gap
$\Delta\propto J^2/\gamma_{\rm loc}$. Thus boundary dissipation can ensure scalable mixing even in the presence of complex bulk dynamics, whereas bulk dissipation generally requires additional control assumptions (see
SM~\cite{SM}).

In the strong-dissipation regime, rapid mixing further requires bounding
$\Tr|\hat\sigma_j|$, as implied by Theorem~\ref{thm:rapid_mixing} and
Eq.~\eqref{eq:doubled_single_prefactor_identity}. Assuming that $\hat K$ is
diagonalizable, we make this bound explicit by evaluating the leading
eigenstate correction in the vectorized $\hat K$-eigenbasis $\{|\epsilon_p,\epsilon_q\rangle\}$
\begin{equation}
|\epsilon_m,\epsilon_n^{(1)}\rangle
=
-\sum_{p}\frac{H_{pm}}{\epsilon^{(0)}_{pm}}\,|\epsilon_p,\epsilon_n^{(0)}\rangle
+\sum_{q}\frac{H_{qn}}{\epsilon^{(0)}_{nq}}\,|\epsilon_m,\epsilon_q^{(0)}\rangle
\end{equation}
where $\{|\epsilon_m\rangle \}$ is the eigenstate of Lindblad operator $\hat{K}$, $H_{pm}\equiv\langle\epsilon_p|\hat{H}|\epsilon_m\rangle$, and
$\epsilon^{(0)}_{pm}$ are the corresponding unperturbed doubled-space energy denominators. Requiring
$\textrm{Tr}|\hat{\sigma}_j|\le \mathrm{poly}(L)$ then translates into a sparsity condition on $\hat{H}$ in the $\{\hat{K}\}$ eigenbasis:
\begin{equation}
\label{trace_norm_require_strong}
\exists\,\alpha\ge \ln{d_0} \ \text{s.t.}\ \forall\,|\epsilon_s\rangle, \ 
N(\alpha,\hat{H},|\epsilon_s\rangle)\le \mathrm{poly}(L).
\end{equation}
Here $N(\alpha,\hat{H},|\epsilon_s\rangle)$ counts the number of indices $p$ for which $|H_{sp}|\ge e^{-\alpha L}$ (with $s$ fixed). See proof in SM \cite{SM}.

\emph{Weak-dissipation regime---} In this regime, the dissipator is treated as a perturbation to the unitary
Schr\"odinger dynamics. The trace-norm prefactor of each eigenmode is therefore controlled by the trace norm of the corresponding right eigenmode, up to small corrections from the perturbatively induced left--right non-orthogonality. We then estimate the mixing-time scaling from two quantities: the Liouvillian gap, set by the leading dissipative correction to the unperturbed spectrum, and the eigenmode trace norm, set by the first-order correction to the eigenvectors.

\textit{Liouvillian gap.}
Consider an unperturbed doubled-space eigenstate $|E_m,E_n^{(0)}\rangle \equiv |E_m\rangle_L\otimes|E_n\rangle_R$ with eigenvalue $\epsilon_{mn}^{(0)}=E_m-E_n$,
where $\{|E_m\rangle\}$ are the eigenstates of $\hat{H}$. The first-order correction is
\begin{equation*}
\langle E_m,E_n^{(0)}|\hat{H}_d|E_m,E_n^{(0)}\rangle
=
-i\gamma\Bigl[\delta K_m^{2}+\delta K_n^{2}
+\bigl(\overline{K_m}-\overline{K_n}\bigr)^{2}\Bigr]
\end{equation*}
with $\overline{K_m}\equiv\langle E_m|\hat{K}|E_m\rangle$ and
$\delta K_m^{2}\equiv \langle E_m|K^{2}|E_m\rangle-\overline{K_m}^{\,2}$.
When $[\hat{K},\hat{H}]\neq0$, typical fluctuations scale as $\delta K_m^{2}\sim L^{a}$ with $a\ge0$. For gapped systems, one expects $a=0$, giving a Liouvillian gap of order $\mathcal{O}(\gamma)$; for gapless systems ($a>0$), enhanced fluctuations can parametrically reduce the gap \footnote{If $[\hat{K},\hat{H}]=0$, then $\delta K_m^{2}=0$ and decay rates are set by the smallest nonzero value of $(\overline{K_m}-\overline{K_n})^{2}$. For local $\hat{K}$ in an ETH-satisfying system, this scale is typically $\mathcal{O}(1)$, yielding a size-independent gap.}. Therefore, a sufficient condition for rapid mixing based on the Liouvillian gap is that $\hat{H}$ is gapped, which ensures a constant gap.

\textit{Trace norm of the eigenmodes.}
The first-order correction to the corresponding eigenmode is
\begin{equation}
\begin{split}
&|E_m,E_n^{(1)}\rangle
=\\
&-i\gamma\sum_{pqst}
\frac{\bigl(K_{ps}-K_{qt}\bigr)\bigl(K_{sm}-K_{tn}\bigr)}
{\epsilon_{mn}^{(0)}-\epsilon_{pq}^{(0)}}
\,|E_p,E_q^{(0)}\rangle
\end{split}
\end{equation}
with $K_{ps}\equiv\langle E_p|\hat{K}|E_s\rangle$. To ensure $\operatorname {Tr}|\hat{\sigma}_j|$ grows at most $\mathrm{poly}(L)$, it suffices to impose a sparsity condition on the matrix elements of $\hat{K}$ in the $\{\ket{E_s}\}$ basis:
\begin{equation}
\label{trace_norm_require_weak}
\exists\,\alpha\ge \ln{d_0}\ \text{s.t.}\ \forall\,\ket{E_s}, \ 
N(\alpha,\hat{K},\ket{E_s})\le \mathrm{poly}(L),
\end{equation}
where $N(\alpha,\hat{K},\ket{E_s})$ counts the number of indices $p$ for which $|\hat{K}_{sp}|\ge e^{-\alpha L}$ (with $s$ fixed). See the SM~\cite{SM} for details. As a simple example, for the nearest-neighbor Ising Hamiltonian (only $\hat{\sigma}_j^{z}\hat{\sigma}_{j+1}^{z}$ terms) with arbitrary single-site Lindblad operators, the energy eigenbasis is a product basis and each local $\hat{K}$ connects only $\mathcal{O}(1)$ configurations; hence $N(\alpha,\hat{K},\ket{E_s})=\mathcal{O}(1)$ (e.g., $N=1$ in the strictly diagonal case).

{\em Discussion.---} In this Letter, using a doubled-space mapping, we directly connect the mixing time to the Liouvillian gap and the trace-norm factors of the decaying eigenmodes of the Liouvillian. This connection provides a clear requirement for achieving fast and rapid mixing. This framework also explains previous findings that mixing time in open systems depends on both the Liouvillian gap and the localization length of the eigenstates \cite{Haga_2021}. Our analysis shows that this is because localization properties directly affect the trace norm of the eigenstates, which in turn influences the mixing time. We analyze the mixing time in both the strong and weak dissipation regimes, where the Liouvillian gap and trace norm can be calculated straightforwardly using perturbation theory. This analysis not only simplifies the problem but also reveals general conditions for fast and rapid mixing.

We stress that our universality refers to the \emph{mixing time}, rather than to the full relaxation dynamics of typical states. Mixing time is a worst-case quantity: it requires \emph{all} initial states to approach the steady state and is therefore controlled by the slowest decaying mode accessible to some initial state. Physically relevant states can relax faster when their overlap with this mode is small or absent. Our framework therefore predicts worst-case mixing while allowing state-dependent faster relaxation.

Our work suggests several directions for future research. First, it would be important to identify general physical conditions under which the trace-norm prefactors of slow Liouvillian eigenmodes remain bounded by $\operatorname{poly}(L)$, and to relate these conditions to locality, dissipation geometry, and correlation length. Second, one should clarify whether changes in the Liouvillian-gap scaling are accompanied by corresponding changes in the trace-norm factor of eigenmodes. For example, by tuning model parameters, one can compare a crossover in the inverse gap, such as from polynomial to logarithmic scaling, with a crossover in the trace-norm factor, such as from exponential to polynomial scaling. Whether these crossovers coincide may help diagnose distinct relaxation mechanisms. A concrete test bed is the boundary-driven transverse-field Ising chain, where both scalings depend on the dissipation strength. Finally, because rapid mixing is essential for experimentally efficient dissipative state preparation, our sparsity constraints may guide the design of feasible dissipative protocols and more efficient Lindblad-state-preparation algorithms.

{\em Acknowledgments.---}
We thank Jerome Lloyd, Tian-Gang Zhou, Pengfei Zhang, Yu-Min Hu, Shuo Liu, and Yuxuan Zhang for helpful discussions. We are grateful to Lorenza Viola and Yikang Zhang for their careful reading of the manuscript and valuable feedback. We also thank Taiki Haga for discussions on related questions.

\appendix

\onecolumngrid

\section{Appendix A: Mixing Time Analysis for Strong Bulk Dissipation}
\label{strong_dissipation_appendix}

In this Appendix, we present a detailed analysis of the mixing time in the strong \emph{bulk}-dissipation regime. In particular, we show that the perturbative approach breaks down for bulk dissipation in the thermodynamic limit.

We begin by specifying the zeroth-order eigenstates and eigenvalues of the doubled-space (vectorized) Hamiltonian. For simplicity, we take the Lindblad operator to be Hermitian. The corresponding zeroth-order generator can then be written as
\begin{equation}
    \hat{H}^{0,D} =-i\gamma \sum_{\alpha=1}^L \left(\hat{K}_{\alpha,L}\otimes \hat{\mathbbm{1}}_R -  \hat{\mathbbm{1}}_L \otimes \hat{K}_{\alpha,R}^* \right)^2.
\end{equation}
The ground state(GS) of the zeroth-order Hamiltonian is the state that satisfies 
\begin{equation}
    \hat{K}_{\alpha,L}|\psi^0_{GS}\rangle=  \hat{K}_{\alpha,R}^{*} |\psi^0_{GS}\rangle
\end{equation}
for each Lindblad operator. If we choose $\{|m\rangle\}$ as a set of complete orthogonal eigenbases for each local site, then the GS can be constructed as
\begin{equation}
    |\psi^0_{GS}\rangle_{\mathbf{m}}^D=\prod_{x=1}^L |m_x\rangle_{x,L} \otimes |m_x\rangle_{x,R}.
    \label{appendix_GS_wavefunction}
\end{equation}
Here, $\mathbf{m}=(m_1,m_2,...,m_L)$ with $m_x = 1,2,...,d_0$, and $d_0$ is the total local Hilbert space dimension. The GS energy is $E_0^{(0)}=0$ and GS degeneracy is $d_0^L$.

We consider the general case that the Hamiltonian does not commute with the Lindblad operators, which means the perturbation can connect the GS to excited states. Since the Hamiltonian acts on all sites, the number of excited states that can be connected to a given ground state by the perturbation is proportional to the total system size $L$. We consider the equal level spacing of local sites, and we denote the level spacing as $1$. For simplicity, we consider the general case where the Hamiltonian does not commute with the jump operators and the Hamiltonian only connects the nearest energy level. Then, the first excited state can be represented as the combination of the below four states (this combination has to make sure that this excited state after the inverse double space mapping gives a tracelss hermian matrix)
\begin{equation}
    |\psi^0_{x_0,e,L\pm}\rangle=|m_{x_0}\pm1\rangle_{x_0,L}\otimes |m_{x_0,R}\rangle_{x_0} \prod_{x\neq x_0} |m_x\rangle_{x,L} \otimes |m_x\rangle_{x,R} ,
\end{equation}
and
\begin{equation}
    |\psi^0_{x_0,e,R\pm}\rangle=|m_{x_0}\rangle_{x_0,L}\otimes |m_{x_0}\pm1\rangle_{x_0,R} \prod_{x\neq x_0} |m_x\rangle_{x,L} \otimes |m_x\rangle_{x,R}. \label{first_order_energy_correction}
\end{equation}
Since $x_0=1,2,...,L$, there are in total $4L$ degenerated first excited states with eigen-energy $E_1^0=-i\gamma$.

Using the degenerated perturbation theory, the first-order correction to the GS energy is
\begin{equation}
    E^{(1)}_{0k}=E^{(0)}_{0k}+V^D_{0k,0k}+\sum_{q} \frac{|V^D_{0k,1q}|^2}{E^{(0)}_{0k}-E^{(0)}_{1q}}.
    \label{appendix_GS_energy_correction}
\end{equation}
Here, $0k$ denotes the label of the zeroth degenerate GS basis set that diagonalizes the $V^D$.  $1q$ denotes the label of $q$-th zeroth excited state. 

We assume that the Hamiltonian does not commute with the Lindblad operators, which means the perturbation can connect the ground state to excited states. Since the Hamiltonian acts on all sites, the number of excited states that can be connected to any given ground state by the perturbation is proportional to the total system size, $L$. We denote the matrix elements of the perturbation as $\langle \psi_{0k}|V^D| \psi_{1q}\rangle=J$, where $|\psi_{0k}\rangle$ represents a GS with label $k$ and $|\psi_{1q}\rangle$ represents a first excited state with label $q$.

Since the smallest diagonal shift $V^{D}_{0k,0k}$ vanishes, the Liouvillian gap is not determined by the first-order (diagonal) correction from $\hat{V}^D $, but instead by the dissipative contribution from the second-order term.

The vanishing of the smallest diagonal shift follows directly from the structure of $\hat{V}^D $. In the vectorized representation,
\begin{equation}
\hat{V}^D = \hat{H} \otimes \hat{\mathbbm{1}}_R -\hat{\mathbbm{1}}_L \otimes \hat{H}^{T}
\end{equation}
so that $\hat{V}^D $ acts as the commutator superoperator,
$\hat{V}^D \,|\rho\rangle\rangle = |[\hat{H},\hat{\rho}]\rangle\rangle$. Hence any operator $\hat{\rho}$ that commutes with $\hat{H}$ is a zero mode of $\hat{V}^D $.
In particular, the identity operator $\hat{\mathbbm{1}}$, any function $f(H)$, and the projectors $\ket{m}\bra{m}$ in the energy eigenbasis satisfy $[\hat{H},\hat{\rho}]=0$, which implies that $\hat{V}^D $ has an exact eigenvalue $\lambda=0$.

By construction, the bi-orthogonal basis $\{\ket{\psi^{R}_{0k}},\bra{\psi^{L}_{0k}}\}$ diagonalizes $\hat{V}^D $ within the zeroth-order degenerate manifold. Therefore, the diagonal matrix elements
\begin{equation}
V^{D}_{0k,0k}\equiv \bra{\psi^{L}_{0k}}\hat{V}^D \ket{\psi^{R}_{0k}}
\end{equation}
coincide with the corresponding eigenvalues of $\hat{V}^D $ restricted to this manifold, and consequently, the smallest value (in magnitude) is
\begin{equation}
\min_{k}|\lambda_k|=0.
\end{equation}
It follows that the non-zero energy correction is controlled solely by the second-order term in Eq.~\eqref{appendix_GS_energy_correction}, yielding the scaling
\begin{equation}
 E^{(1)}_{0k} \propto -\, i\,L\,\frac{J^{2}}{\gamma}.
\end{equation}

Because this energy correction diverges as $L \to \infty$, this perturbation analysis is not applicable in the thermodynamic limit.

We emphasize that the breakdown of the above perturbative estimate should not be interpreted as evidence that rapid mixing is impossible in the strong bulk-dissipation regime. The quantity of order $LJ^2/\gamma$ is an extensive total escape rate from a fixed product configuration, such as the one in Eq.~(3). It arises from $O(L)$ local virtual transitions to fast-decaying excited states and therefore should not be identified with the Liouvillian gap.

To determine the gap, one must instead treat the full degenerate zero-mode manifold of $\hat{\mathcal{L}}_0$. Equivalently, one can eliminate the fast-decaying coherences perturbatively. This gives an effective second-order generator acting within the slow Zeno manifold,
\begin{equation}
    \hat{\mathcal{L}}_{\rm eff}^{(2)}
    =
    -\hat{P}\hat{\mathcal{V}}\hat{Q}
    \left(\hat{Q}\hat{\mathcal{L}}_0 \hat{Q} \right)^{-1}
    \hat{Q} \hat{\mathcal{V}}\hat{P} ,
\end{equation}
where $\hat{P}$ projects onto the kernel of $\hat{\mathcal{L}}_0$, $\hat{Q}=\hat{\mathbbm{1}}-\hat{P}$, and $\hat{\mathcal{V}}=-i[\hat{H},\cdot]$. In the basis that diagonalizes the local jump operators, this effective generator acts on diagonal density matrices as a classical Markov generator, with transition rates of order $J^2/\gamma$. Thus, the presence or absence of rapid mixing in this regime is controlled by the spectral properties of the induced stochastic process, rather than by the naive extensive escape rate. These properties are model dependent: the effective generator may have a finite gap, a diffusive gap, or a gap controlled by kinetic constraints or conservation laws. Consequently, no universal scaling of the Liouvillian gap can be inferred from the naive perturbative estimate in the strong bulk-dissipation regime. In particular, rapid mixing may still occur if the effective process is sufficiently connected and has no slow conserved modes, whereas diffusive or constrained effective dynamics may lead to a gap that closes with system size. Thus, the strong bulk-dissipation regime can host alternative rapid-mixing mechanisms, but these mechanisms lie outside the controlled perturbative criterion discussed in the main text.

\section{Appendix B: Constraints on the Decaying Eigenmodes in the Strong- and Weak-dissipation Regimes}
\label{appendix_rapid_mix_constraints}

In this appendix, we establish sufficient conditions under which the trace norm of each decaying right eigenmode satisfies $\Tr|\hat{\sigma}_j|
    =
    \|\hat{\sigma}_j\|_1
    \leq$ in both the strong- and weak-dissipation regimes, as stated in
Eqs.~\eqref{trace_norm_require_strong} and \eqref{trace_norm_require_weak}. This condition is directly related to the trace-norm requirement for rapid mixing in Theorem~2 of the main text. There, the required bound is imposed on the trace-norm prefactor $C_j=\|\hat P_j\|_{1\to 1}$ of each decaying spectral projector. In the strong- and weak-dissipation regimes considered here, the left and right eigenmodes remain close to being orthogonal, up to perturbative corrections. Consequently, the projector prefactor can be bounded by the trace norm of the corresponding right eigenmode, schematically as $C_j
    \lesssim
    \|\hat{\sigma}_j\|_1^2$, up to small corrections due to the weak non-orthogonality between left and right eigenmodes.

Therefore, in these perturbative regimes, it is sufficient to show that the right eigenmodes have at most polynomial trace norm in system size. This implies that
\begin{equation}
    \log C_j
    \leq
    \mathcal O\!\left[\operatorname{poly}(\log L)\right],
\end{equation}
which is the trace-norm condition required for rapid mixing.

Throughout, the total Hilbert-space dimension is $N=d^L$. We use $\|\cdot\|_1$ for the trace norm and $\|\cdot\|_2$ for
the Frobenius norm, $\|\hat{X}\|_2^2=\mathrm{Tr}(\hat{X}^\dagger \hat{X})$. For the lowest excited state operator $\hat{\sigma}_1$ we assume the
normalization $\mathrm{Tr}(\hat{\sigma}_1^2)=1$ (Hermitian $\hat{\sigma}_1$), which implies $\|\hat{\sigma}_1\|_2=1$.

% ============================================================
\subsection{Strong-dissipation regime}
\label{app:strong_constraints}

In the strong-dissipation regime, we assume that $\hat{K}$ is diagonalizable. The unperturbed doubled-space (vectorized) eigenbasis is
$\{|\epsilon_p,\epsilon_q\rangle\}\equiv \{|\epsilon_p\rangle_L\otimes |\epsilon_q\rangle_R\}$,
where $|\epsilon_s\rangle$ are simultaneous eigenstates of the Lindblad operators $\{\hat{K}_j\}$.
The leading correction to the state $|\epsilon_m,\epsilon_m\rangle$, induced by the Hamiltonian superoperator
\begin{equation}
\hat{H}_s \;=\; \hat{H}_L\otimes \hat{\mathbbm{1}}_R \;-\; \hat{\mathbbm{1}}_L\otimes \hat{H}_R^{T}
\end{equation}
takes the standard first-order perturbative form
\begin{equation}
    \begin{split}
    |\epsilon_m,\epsilon_m^{(1)}\rangle
    =\sum_{p,q}\frac{\langle \epsilon_p,\epsilon_q|\hat{H}_s|\epsilon_m,\epsilon_m\rangle}{\epsilon^{(0)}_{mm}-\epsilon^{(0)}_{pq}}\,|\epsilon_p,\epsilon_q\rangle =
    -\sum_{p}\frac{H_{pm}}{\epsilon^{(0)}_{pm}}\,|\epsilon_p,\epsilon_m^{(0)}\rangle
    +\sum_{q}\frac{H_{qm}}{\epsilon^{(0)}_{mq}}\,|\epsilon_m,\epsilon_q^{(0)}\rangle ,
    \end{split}
\label{eq:strong_first_order_state}
\end{equation}
where $H_{pm}\equiv \langle \epsilon_p|\hat{H}|\epsilon_m\rangle$, and $\epsilon^{(0)}_{pq}$ denotes the unperturbed doubled-space eigenvalue associated with the basis state $|\epsilon_p,\epsilon_q\rangle$.
Here, $\langle \epsilon_p|$ denotes the $p$th left eigenvector of $\hat{K}$. For a general non-Hermitian $\hat{K}$, the left eigenstate is not necessarily the Hermitian (complex-conjugate) transpose of the corresponding right eigenstate $|\epsilon_p\rangle$.

It is convenient to write the perturbed vectorized state as
\begin{equation}
|\epsilon_m,\epsilon_m^{(1)}\rangle=\sum_{p,q} C_{pq}\,|\epsilon_p,\epsilon_q^{(0)}\rangle,
\label{eq:Cpq_def_strong}
\end{equation}
which corresponds to an operator (matrix) representation
\begin{equation}
\hat{\sigma}_1=\sum_{p,q} C_{pq}\,|\epsilon_p\rangle\langle \epsilon_q|.
\label{eq:sigma1_from_Cpq}
\end{equation}
From Eq.~\eqref{eq:strong_first_order_state}, the coefficients are
\begin{equation}
C_{pq}
=\frac{H_{qm}}{\epsilon^{(0)}_{mq}}\delta_{pm}
-\frac{H_{pm}}{\epsilon^{(0)}_{pm}}\delta_{qm},
\qquad (p\neq q).
\label{eq:C_element}
\end{equation}
Equation~\eqref{eq:C_element} shows that $\hat{\sigma}_1$ has at most $2N$ nonzero matrix elements in the $\{|E_s\rangle\}$
basis: only the $m$-th row and $m$-th column can be populated.

\paragraph{Cutoff decomposition.}
Fix a cutoff $c>0$. Let $\hat{A}$ be the matrix obtained from $\hat{\sigma}_1$ by keeping only those entries with magnitude larger
than $c$, and let $\hat{B}:=\hat{\sigma}_1-\hat{A}$ be the remainder. Thus $\hat{\sigma}_1=\hat{A}+ \hat{B}$ and by the triangle inequality
\begin{equation}
\|\hat{\sigma}_1\|_1 \le \|\hat{A}\|_1+\|\hat{B}\|_1.
\label{eq:triangle_AB}
\end{equation}
We bound $\|\hat{A}\|_1$ and $\|\hat{B}\|_1$ separately.

\paragraph{Bound on the ``large'' part $\hat{A}$.}
Using $\|\hat{X}\|_1\le \sqrt{\mathrm{rank}(\hat{X})}\,\|\hat{X}\|_2$ and $\|\hat{A}\|_2\le \|\hat{\sigma}_1\|_2=1$,
\begin{equation}
\|\hat{A}\|_1 \le \sqrt{\mathrm{rank}(\hat{A})}\,\|\hat{A}\|_2 \le \sqrt{\mathrm{rank}(\hat{A})} \le \sqrt{N_{\mathrm{large}}},
\label{eq:A_bound_strong}
\end{equation}
where $N_{\mathrm{large}}$ is the number of entries of $\hat{\sigma}_1$ with magnitude $>c$, and we used $\mathrm{rank}(\hat{A})\le
N_{\mathrm{large}}$. Therefore, a sufficient condition for $N_{\rm{band}}\|\hat{A}\|_1\le \mathrm{poly}(L)$ is
\begin{equation}
N_{\mathrm{large}} \le \mathrm{poly}(L).
\label{eq:Nlarge_poly_condition}
\end{equation}

\paragraph{Bound on the ``small'' part $B$.}
Again using $\|\hat{X}\|_1\le \sqrt{\mathrm{rank}(X)}\,\|\hat{X}\|_2$ and $\mathrm{rank}(\hat{B})\le N$, we get
\begin{equation}
\|\hat{B}\|_1 \le \sqrt{N}\,\|\hat{B}\|_2.
\label{eq:B_rank_bound}
\end{equation}
Moreover, since $\hat{B}$ contains only entries of magnitude $\le c$ and $\hat{\sigma}_1$ has at most $2N$ nonzero entries in this basis, we have the Frobenius bound
\begin{equation}
\|\hat{B}\|_2^2=\sum_{i,j}|B_{ij}|^2 \le (2N)\,c^2,
\qquad\Rightarrow\qquad
\|\hat{B}\|_1 \le \sqrt{N}\,\sqrt{2N}\,c \;=\;\sqrt{2}\,N\,c.
\label{eq:B_bound_strong}
\end{equation}
Hence a sufficient condition for $\|\hat{B}\|_1\le \mathrm{poly}(L)$ is
\begin{equation}
c \le \frac{\mathrm{poly}(L)}{N}.
\label{eq:c_condition_strong}
\end{equation}

\paragraph{From cutoff bounds to a sparsity condition on $\hat{H}$.}
From Eq.~\eqref{eq:C_element}, the only potentially nonzero entries of $\hat{\sigma}_1$ are proportional to $|H_{pm}|/|\epsilon_{pm}^{(0)}|$ and $|H_{qm}|/|\epsilon_{mq}^{(0)}|$. In the strong-dissipation regime, the denominators are set by the dissipative scale and remain bounded below by an $L$-independent constant (up to model-dependent factors), so bounding entries of $\hat{\sigma}_1$ reduces to bounding the matrix elements of $\hat{H}$ in the $\{|\epsilon_s\rangle\}$ basis. Choosing
a cutoff that scales as $c\sim e^{-\alpha L}$ leads to the following sufficient condition:
\begin{equation}
\label{trace_norm_require_strong}
\exists\,\alpha\ge \ln d_0\ \text{s.t.}\ \forall\,|\epsilon_s\rangle,\qquad
N(\alpha,\hat{H},|\epsilon_s\rangle)\le \mathrm{poly}(L),
\end{equation}
where $N(\alpha,\hat{H},|\epsilon_s\rangle)$ counts the number of indices $p$ such that
\begin{equation}
|H_{sp}| \ge e^{-\alpha L}
\qquad\text{for a fixed eigenstate }|\epsilon_s\rangle\text{ of }\{\hat{K}_j\}.
\end{equation}
Together with Eqs.~\eqref{eq:Nlarge_poly_condition} and \eqref{eq:c_condition_strong}, this implies
$\|\hat{\sigma}_1\|_1=\mathrm{Tr}|\hat{\sigma}_1|\le \mathrm{poly}(L)$. Here, $d_0$ is the local Hilbert-space dimension, so the total Hilbert-space dimension is $N=d_0^L=e^{(\ln d_0)L}$. The cutoff $c$ is chosen to decay exponentially because it must satisfy $c \lesssim \mathrm{poly}(L)\,N^{-1}$. Since $N^{-1}$ itself decays exponentially with $L$, slower scalings such as $c\sim \mathrm{poly}(L)^{-1}$ are generally insufficient.

\paragraph{Other excited states.}
For arbitrary excited states, the first-order correction can be written as 
\begin{equation}
\begin{split}
|\epsilon_m,\epsilon_n^{(1)}\rangle
&=\sum_{p,q}\frac{\langle \epsilon_p,\epsilon_q|\hat{H}_s|\epsilon_m,\epsilon_n\rangle}{\epsilon^{(0)}_{mn}-\epsilon^{(0)}_{pq}}\,|\epsilon_p,\epsilon_q\rangle \\
&=
\sum_{p}\frac{H_{pm}}{\epsilon^{(0)}_{mn}-\epsilon^{(0)}_{pn}}\,|\epsilon_p,\epsilon_n^{(0)}\rangle
+\sum_{q}\frac{H_{qn}}{\epsilon^{(0)}_{mn}-\epsilon^{(0)}_{mq}}\,|\epsilon_m,\epsilon_q^{(0)}\rangle.
\end{split}
\end{equation}
Following a similar analysis, we find that the sufficient condition for the trace norm of an arbitrary excited state to be bounded by $\mathrm{poly}(L)$ is identical to that of the lowest excited state. Thus, the sparsity constraints for rapid mixing apply universally across the entire spectrum, rather than being restricted to the lowest-lying states.

\subsection{Weak-dissipation regime}
\label{app:weak_constraints}

In the weak-dissipation regime, the leading correction to doubled-space eigenstates is generated by the dissipator. Let $\{|E_p,E_q\rangle\}$ be the doubled-space basis built from energy eigenstates $\{|E_s\rangle\}$ of the Hamiltonian, and let $|E_m,E_n^{(0)}\rangle$ denote an unperturbed eigenstate with eigenvalue $\epsilon_{mn}^{(0)}$. To first order in the dissipation strength $\gamma$, standard (non-degenerate) perturbation theory gives
\begin{equation}
\begin{split}
|E_m,E_n^{(1)}\rangle
&= -i\gamma\sum_{p,q}
\frac{\langle E_p,E_q^{(0)}|\left(\hat{K}_L-\hat{K}_R^{*}\right)^2|E_m,E_n^{(0)}\rangle}
{\epsilon_{mn}^{(0)}-\epsilon_{pq}^{(0)}}
\,|E_p,E_q^{(0)}\rangle \\
&= -i\gamma\sum_{p}
\frac{\langle E_p|\hat{K}^2|E_m\rangle}{\epsilon_{mn}^{(0)}-\epsilon_{pn}^{(0)}}
\,|E_p,E_n^{(0)}\rangle
-i\gamma\sum_{q}
\frac{\langle E_q|\hat{K}^2|E_n\rangle}{\epsilon_{mn}^{(0)}-\epsilon_{mq}^{(0)}}
\,|E_m,E_q^{(0)}\rangle \\
&\quad
+2i\gamma\sum_{p,q}
\frac{\langle E_p|\hat{K}|E_m\rangle\,\langle E_q|\hat{K}|E_n\rangle}
{\epsilon_{mn}^{(0)}-\epsilon_{pq}^{(0)}}
\,|E_p,E_q^{(0)}\rangle .
\end{split}
\label{eq:C_weak}
\end{equation}
Here $\hat{K}_L$ and $\hat{K}_R$ denote left- and right-multiplication by $\hat{K}$ in the doubled space, and we use the shorthand
$K_{pm}\equiv \langle E_p|\hat{K}|E_m\rangle$.

As in the strong-dissipation analysis, we unvectorize the perturbed doubled-space vector and view it as an operator
$\hat{\sigma}_1$, whose matrix elements are precisely the coefficients in Eq.~\eqref{eq:C_weak}.
We bound $\|\hat{\sigma}_1\|_1$ by splitting
\begin{equation}
\hat{\sigma}_1 = \hat{A} + \hat{B}
\end{equation}
with a cutoff $c>0$: entries with magnitude $\ge c$ are assigned to $\hat{A}$ (the ``large'' part), and the remaining entries to $\hat{B}$ (the ``small'' part). By the triangle inequality \eqref{eq:triangle_AB}, it suffices to control $\|\hat{A}\|_1$ and
$\|\hat{B}\|_1$ separately.

\paragraph{Large part $\hat{A}$.}
The same counting argument as in the strong-dissipation case yields
\begin{equation}
\|\hat{A}\|_1 \le \sqrt{N_{\mathrm{large}}},
\end{equation}
where $N_{\mathrm{large}}$ is the number of matrix elements of $\hat{\sigma}_1$ whose magnitude is at least $c$.
Hence $\|\hat{A}\|_1\le \mathrm{poly}(L)$ follows from $N_{\mathrm{large}}\le \mathrm{poly}(L)$.

\paragraph{Small part $\hat{B}$.}
Equation~\eqref{eq:C_weak} contains two structurally distinct sets of coefficients:
\begin{enumerate}
\item terms proportional to $K_{pm}K_{qn}$ (the double sum in the last line), which populate $O(N^2)$ entries in the
$\{|E_p,E_q\rangle\}$ basis;
\item terms proportional to $(K^2)_{pm}$ or $(K^2)_{qn}$ (the first two sums), which populate $O(N)$ entries.
\end{enumerate}
Accordingly, we decompose
\begin{equation}
\hat{B} = \hat{B}_1 + \hat{B}_2,
\end{equation}
where $\hat{B}_1$ collects the $O(N^2)$ coefficients proportional to $K_{pm}K_{qn}$ and $\hat{B}_2$ collects the $O(N)$ coefficients
proportional to $(K^2)_{pm}$ or $(K^2)_{qn}$.

By construction, every entry of $\hat{B}$ has magnitude at most $c$. Therefore,
\begin{equation}
\|\hat{B}_1\|_2^2 \le N^2 c^2
\qquad\Rightarrow\qquad
\|\hat{B}_1\|_1 \le \sqrt{N}\,\|\hat{B}_1\|_2 \le N^{3/2}c,
\label{eq:B1_bound_weak}
\end{equation}
so $\|\hat{B}_1\|_1\le \mathrm{poly}(L)$ is ensured by
\begin{equation}
c \le \frac{\mathrm{poly}(L)}{N^{3/2}}.
\label{eq:c_condition_weak_B1}
\end{equation}
Similarly,
\begin{equation}
\|\hat{B}_2\|_2^2 \le N c^2
\qquad\Rightarrow\qquad
\|\hat{B}_2\|_1 \le \sqrt{N}\,\|\hat{B}_2\|_2 \le Nc,
\label{eq:B2_bound_weak}
\end{equation}
and $\|\hat{B}_2\|_1\le \mathrm{poly}(L)$ follows from
\begin{equation}
c \le \frac{\mathrm{poly}(L)}{N}.
\label{eq:c_condition_weak_B2}
\end{equation}

\paragraph{From cutoff bounds to conditions on matrix elements of $\hat{K}$.}
To translate the cutoff constraints into a condition on $\hat{K}$, we relate the typical size of the coefficients in $\hat{B}_1$
and $\hat{B}_2$ to the matrix elements $K_{pm}$.

For the $\hat{B}_1$ sector, the relevant coefficients scale as products $K_{pm}K_{qn}$; hence a conservative estimate is
\begin{equation}
|K_{pm}K_{qn}| \lesssim c
\qquad\Rightarrow\qquad
\max_{p,m}|K_{pm}| \lesssim \sqrt{c}.
\end{equation}
Combining with Eq.~\eqref{eq:c_condition_weak_B1} yields the sufficient requirement
\begin{equation}
\max_{p,m}|K_{pm}| \;\lesssim\; \frac{\mathrm{poly}(L)}{N^{3/4}}.
\end{equation}

For the $\hat{B}_2$ sector, coefficients involve $(K^2)_{pm}=\sum_{r}K_{pr}K_{rm}$. Using the bound
\begin{equation}
|(K^2)_{pm}| \le \sum_{r}|K_{pr}||K_{rm}|
\le N\big(\max_{a,b}|K_{ab}|\big)^2,
\end{equation}
we obtain
\begin{equation}
|(K^2)_{pm}| \lesssim c
\qquad\Rightarrow\qquad
\max_{p,m}|K_{pm}| \lesssim \sqrt{\frac{c}{N}}.
\end{equation}
Combining with Eq.~\eqref{eq:c_condition_weak_B2} gives the sufficient requirement
\begin{equation}
\max_{p,m}|K_{pm}|
\;\lesssim\;
\frac{\mathrm{poly}(L)}{N}.
\end{equation}
Since this is stronger than the $\hat{B}_1$ condition for large $N$, it is sufficient to impose
\begin{equation}
\max_{p,m}|K_{pm}|\;\lesssim\;\frac{\mathrm{poly}(L)}{N}.
\end{equation}

Equivalently, choosing an exponential cutoff $c\sim e^{-\alpha L}$ turns the above bounds into a sparsity requirement
on $\hat{K}$ in the energy eigenbasis. A sufficient condition is that there exists $\alpha\ge \ln{d_0}$ such that, for every
eigenstate $|E_s\rangle$, only polynomially many matrix elements exceed the cutoff:
\begin{equation}
\label{trace_norm_require_weak}
\exists\,\alpha\ge \ln{d_0}\ \text{s.t.}\ \forall\,|E_s\rangle,\qquad
N(\alpha,\hat{K},|E_s\rangle)\le \mathrm{poly}(L),
\end{equation}
where $N(\alpha,\hat{K},|E_s\rangle)$ counts the number of indices $p$ such that
\begin{equation}
|K_{sp}| \ge e^{-\alpha L}.
\end{equation}

Combining the bounds on $\hat{A}$ and $\hat{B}$ with Eq.~\eqref{trace_norm_require_weak} and the cutoff choice $c\sim e^{-\alpha L}$
yields
\begin{equation}
\|\hat{\sigma}_1\|_1=\mathrm{Tr}|\hat{\sigma}_1|\le \mathrm{poly}(L)
\end{equation}
in the weak-dissipation regime.

\section{Appendix C: Two Equivalent Ways of Rewriting the Mixing Time}
\label{appendix_two_way_definition}

In this appendix, we clarify two equivalent conventions for rewriting the mixing time.
The first is the convention adopted in the main text, while the second is commonly used in earlier works.
The apparent discrepancy between the two is purely a matter of normalization (i.e., where one ``stores'' an extensive factor), and does not reflect any physical inconsistency.

One example is the statement in Ref.~\cite{PhysRevResearch.3.043137} that the prefactor of the slowest decaying mode scales as $e^{L}$, whereas in our main text, when the initial state simply chosen as: $|\psi_{\rho_0}^D\rangle
    =
    |\psi_{\sigma_0}\rangle
    +
    c_1|\psi_{\sigma_1}\rangle$,  the coefficient $c_1$ is at most $O(1)$.
These statements are consistent because the product of the coefficient and the mode---which is what enters the physical trace distance---is invariant under rescaling of eigenmodes.

\subsection{Way 1 (main text): Hilbert--Schmidt normalization}

In the main text, we normalize the slowest decaying right mode by its Hilbert--Schmidt norm,
\begin{equation}
    \Tr(\hat{\sigma}_1^\dagger \hat{\sigma}_1)=1 .
\end{equation}
For Hermitian modes, this reduces to $\Tr(\hat{\sigma}_1^2)=1$. Consider an initial state whose deviation from the steady state is along this slowest mode,
\begin{equation}
    \hat{\rho}_0=\hat{\sigma}_0+c_1\hat{\sigma}_1 .
\end{equation}
Since both $\hat{\rho}_0$ and $\hat{\sigma}_0$ are physical density matrices, their Hilbert--Schmidt norms satisfy
\begin{equation}
    \|\hat{\rho}_0\|_2=\sqrt{\Tr(\hat{\rho}_0^2)}\le 1,
    \qquad
    \|\hat{\sigma}_0\|_2=\sqrt{\Tr(\hat{\sigma}_0^2)}\le 1 .
\end{equation}
Using the triangle inequality, we obtain
\begin{equation}
    |c_1|
    =
    \|c_1\hat{\sigma}_1\|_2
    =
    \|\hat{\rho}_0-\hat{\sigma}_0\|_2
    \le
    \|\hat{\rho}_0\|_2+\|\hat{\sigma}_0\|_2
    \le 2 .
\end{equation}
Thus, under this normalization, the overlap coefficient $c_1$ is bounded by an $O(1)$ number and cannot grow with system size.
With this convention, any possible system-size enhancement in the contribution of the slowest mode cannot come from $c_1$. It must instead come from the trace norm $\Tr|\hat{\sigma}_1|$ appearing in the trace distance.

\subsection{Way 2 (common in the literature): trace-norm normalization}

Some references, for example \cite{PhysRevResearch.3.043137,Haga_2021} instead normalize the slow mode so that its trace norm is $O(1)$, e.g.
\begin{equation}
\Tr|\tilde{\sigma}_1|=1,
\end{equation}
which can be achieved by rescaling
\begin{equation}
\tilde{\sigma}_1 = a\,\hat{\sigma}_1,
\qquad
\tilde{c}_1 = c_1/a.
\end{equation}
If $\Tr|\hat{\sigma}_1|\sim e^{L}$ in our convention, then choosing $a\sim 1/\Tr|\hat{\sigma}_1|$ makes $\Tr|\tilde{\sigma}_1|=O(1)$ and transfers the exponential factor into $\tilde{c}_1$.
This is exactly how one can obtain an $e^{L}$-scaling coefficient in Ref.~\cite{PhysRevResearch.3.043137} while still being consistent with $c_1=O(1)$ in our convention.

\subsection{Invariant quantity and mixing time}

The trace distance to the steady state depends only on the product of the coefficient and the trace norm of the mode.
Keeping only the slowest mode for simplicity, the time-evolved contribution reads
\begin{equation}
\hat{\rho}(t)-\hat{\sigma}_0 =c_1\,e^{-i\alpha_1 t-\Delta t}\,\hat{\sigma}_1,
\end{equation}
and hence
\begin{equation}
D(\hat{\rho}(t),\hat{\sigma}_0)
=\frac{1}{2}\Tr|\hat{\rho}(t)-\hat{\sigma}_0|
= \frac{|c_1|}{2}\,e^{-\Delta t}\,\Tr|\hat{\sigma}_1|.
\label{eq:trace_distance_product}
\end{equation}
Under the rescaling $\hat{\sigma}_1\mapsto \tilde{\sigma}_1=a\hat{\sigma}_1$ and $c_1\mapsto \tilde{c}_1=c_1/a$, the product $|c_1|\Tr|\hat{\sigma}_1|$ is unchanged, and therefore the mixing time extracted from \eqref{eq:trace_distance_product} is identical in both conventions. In other words, different normalizations merely redistribute the same size dependence between the coefficient and the eigen-operator, without changing any physical prediction.

\section{Appendix D: a stronger but simpler sufficient condition for fast mixing and rapid mixing }
A stronger but simpler sufficient condition for the rapid mixing or fast mixing is that the full doubled-space eigenbasis is upper bounded and Liouvillian gap is also upper bounded. Writing
\begin{equation}
    \hat H^D=\hat{V} \hat{E} \hat{V}^{-1},
\end{equation}
we define the trace-norm condition number
\begin{equation}
    \kappa_1(\hat{V})
    =
    \|\hat{V}\|_{1\to 1}
    \|\hat{V}^{-1}\|_{1\to 1},
    \label{eq:doubled_condition_number_def}
\end{equation}
where the induced norm is understood after unvectorization, as the spectrum projector factor defined as
\begin{equation}
    C_j =
    \|\hat P_j\|_{1\to 1}
    \equiv
    \sup_{\|\hat X\|_1=1}
    \|\hat P_j (\hat X )\|_1,
    \label{eq:doubled_band_prefactor_def}
\end{equation}
if $\kappa_1(\hat{V})\leq \operatorname{poly}(L)$, then all spectral projectors, and hence all eigenmode projectors, are
polynomially bounded in trace norm:
\begin{equation}
    C_j\leq \operatorname{poly}(L).
\end{equation}
Equation
\begin{equation}
    \tau_{\rm mix}(\eta)
    \lesssim
    \max_j
    \frac{1}{\Gamma_j}
    \log\left(
    \frac{C_j}{\eta}
    \right).
    \label{eq:doubled_band_mixing_time_bound}
\end{equation}
then reduces to
\begin{equation}
    \tau_{\rm mix}(\eta)
    \lesssim
    \max_j
    \frac{1}{\Gamma_j}
    \log\left(
    \frac{\operatorname{poly}(L)}{\eta}
    \right).
\end{equation}
If the trace-norm condition number satisfies $\kappa_1(\hat{V})\leq \mathcal O[\operatorname{poly}( L)]$, and the inverse Liouvillian gap is also bounded as $\Delta^{-1}\leq \mathcal O[\operatorname{poly}(\log L)]$, then, since $\Gamma_j\geq \Delta$, then for all decaying eigenmodes,
\begin{equation}
    \Gamma_j^{-1}\leq \Delta^{-1}
    \leq \mathcal O[\operatorname{poly}(\log L)] .
\end{equation}
Combining this bound with Eq.~\eqref{eq:doubled_band_mixing_time_bound},
we obtain
\begin{equation}
    \tau_{\rm mix}(\eta)
    \leq
    \mathcal O[\operatorname{poly}(\log L)] ,
\end{equation}
for fixed accuracy $\eta$. Thus the dynamics is rapidly mixing. 

Similarly, if $\kappa_1( \hat{V} ) \leq O[\operatorname{poly}(L)]$ and $\Delta^{-1}\leq \mathcal O[\operatorname{poly}(L)]$, then Eq.~\eqref{eq:doubled_band_mixing_time_bound} gives
\begin{equation}
    \tau_{\rm mix}(\eta)
    \leq
    \mathcal O[\operatorname{poly}(L)] 
\end{equation}
 for fixed $\eta$. Hence the dynamics is fast mixing.

This global conditioning assumption is sufficient but stronger than
necessary. The essential requirement for mixing is the weighted
spectral-sector bound in Eq.~\eqref{eq:doubled_band_mixing_time_bound}:
a rapidly decaying sector may have a large projector prefactor without
dominating the mixing time.

\section{Appendix E: Role of the Number of Decaying Eigenmodes in the Rapid-mixing Criterion}
\label{app:Nband}

In this appendix, we discuss why the number of decaying eigemmodes $N_{\rm{mode}}$ is not expected to modify the rapid-mixing
scaling in the generic situations considered in this work. The discussion
below is not intended as a fully general theorem for arbitrary non-normal
Liouvillians. Rather, it explains why the possible $N_{\rm band}$ factor
arising from a crude triangle-inequality estimate is usually not the
scaling-limiting contribution.

Let $\delta\hat{\rho}(t)
    =
    \hat{\rho}(t)-\hat{\sigma}_0$ be the deviation from the steady state. Decomposing the dynamics into
decaying eigenmodes gives 
$$
    \delta\hat{\rho}(t)
    =
    \sum_j \delta \hat{\rho}_j(t),
    \qquad
    \delta \hat{\rho}_j(t)
    =
    e^{t\hat{\mathcal{L}}} \hat{P}_j \delta \hat{\rho}(0),
$$
where $\hat{P}_j$ is the spectral projection onto the decay eigenmode $\hat{\sigma}_j$. If the decay rate of this eigenmode is $\Gamma_j$, then
$$
    \|\delta \hat{\rho}_j(t)\|_1
    \lesssim
    C_j e^{-\Gamma_j t}\|\delta \hat{\rho}(0)\|_1 ,
$$
where $C_j=\|\hat{P}_j\|_{1\rightarrow 1}$ is the trace-norm factor associated with the spectral projection. Therefore, a crude estimate gives
\begin{equation}
     D(\hat{\rho}(t),\hat{\sigma}_0)
    =
    \frac12\left\|\sum_j \delta \hat{\rho}_j(t)\right\|_1
    \leq
    \frac12\sum_j \|\delta \hat{\rho}_j(t)\|_1 .
\end{equation}
If we require each of the decaying eigenmode contribution to be smaller than $\eta$, this
bound gives
\begin{equation}
     D(\hat{\rho}(t),\hat{\sigma}_0)
    \leq
    N_{\rm{mode}}\eta .
\end{equation}
Thus, a completely rigorous bound based only on the triangle inequality would
require each band to be smaller than $\eta/N_{\rm{mode}}$, producing an
additional factor $\log N_{\rm{mode}}$
in the mixing-time estimate. If $N_{\rm{mode}}$ were exponentially large in
the system size, this factor would be potentially dangerous for rapid mixing.

However, this $N_{\rm{mode}}$ dependence is a worst-case artifact of the
triangle inequality. Saturating the bound requires many different decaying
bands to contribute to the trace norm with essentially no cancellation.
Generically, different Liouvillian eigenmodes have different singular-vector
structures and, when the eigenvalues are complex, different phases under time
evolution. Hence the sum $\sum_j \delta \hat{\rho}_j(t)$ does not usually add coherently in trace norm. The triangle-inequality bound therefore substantially overestimates the actual trace distance.

The most dangerous case is when many decaying contributions add without
cancellation. A clean model of this situation is the case where these
contributions have mutually orthogonal supports. We now show that even in
this exceptional case the additional relaxation time is controlled by the
Liouvillian gap, rather than by the number of such modes.

\begin{lemma}[Decay of mutually support-orthogonal modes]
Let
$$
    \delta \hat{\rho}_{\perp}(t)
    =
    \sum_{j\in\mathcal S} \hat{X}_j(t),
    \qquad
    \hat{X}_j(t)
    =
    c_j e^{\lambda_j t}\hat{\sigma}_j ,
$$
where $ \lambda_j=-\Gamma_j+i\omega_j, \ 
    \Gamma_j\geq \Delta>0$. Assume that the operators $\hat{X}_j(t)$ are mutually support-orthogonal, namely
there exist orthogonal projectors $\hat{P}_j$ such that
$$
    \hat{P}_j\hat{P}_k=0
    \quad (j\neq k),
    \qquad
    \hat{X}_j(t)=\hat{P}_j\hat{X}_j(t)\hat{P}_j .
$$
Then, $\left\|\delta \hat{\rho}_{\perp}(t)\right\|_1
    =
    \sum_{j\in\mathcal S}\|\hat{X}_j(t)\|_1$.

Moreover, after an additional time $T_{\rm add}$,
$$ \left\|\delta \hat{\rho}_{\perp}(t+T_{\rm add})\right\|_1
    \leq
    e^{-\Delta T_{\rm add}}
    \left\|\delta \hat{\rho}_{\perp}(t)\right\|_1 .$$
Equivalently,
$$
    D_{\perp}(t+T_{\rm add})
    \leq
    e^{-\Delta T_{\rm add}}D_{\perp}(t),
$$
where $D_{\perp}(t)
    =
    \frac12\|\delta \hat{\rho}_{\perp}(t)\|_1$.
\end{lemma}

\begin{proof}
Because the operators $\hat{X}_j(t)$ are supported on mutually orthogonal blocks,
the singular values of their sum are simply the union of the singular values of the individual blocks. Hence
\begin{equation}
    \left\|\sum_{j\in\mathcal S}\hat{X}_j(t)\right\|_1
    =
    \sum_{j\in\mathcal S}\|\hat{X}_j(t)\|_1 .
\end{equation}
Furthermore,
\begin{equation}
     \hat{X}_j(t+T_{\rm add})
    =
    e^{\lambda_j T_{\rm add}}\hat{X}_j(t),
\end{equation}
and therefore
$$
    \|\hat{X}_j(t+T_{\rm add})\|_1
    =
    e^{-\Gamma_j T_{\rm add}}\|\hat{X}_j(t)\|_1
    \leq
    e^{-\Delta T_{\rm add}}\|\hat{X}_j(t)\|_1 .
$$
Using the trace-norm additivity over orthogonal blocks, we obtain
\begin{equation}
    \left\|\delta \hat{\rho}_{\perp}(t+T_{\rm add})\right\|_1 =
    \sum_{j\in\mathcal S}\|\hat{X}_j(t+T_{\rm add})\|_1 \leq
    e^{-\Delta T_{\rm add}}
    \sum_{j\in\mathcal S}\|\hat{X}_j(t)\|_1
    =
    e^{-\Delta T_{\rm add}}
    \left\|\delta \hat{\rho}_{\perp}(t)\right\|_1.
\end{equation}
Dividing by $2$ gives the corresponding statement for the trace distance.
\end{proof}

This lemma shows that even when many decaying modes add directly in trace norm, their combined contribution decays as a whole with the Liouvillian gap. Indeed, if $\delta \hat{\rho}_{\perp}(t)$ is an uncancelled block-direct-sum component of the physical deviation $\delta\hat{\rho}(t)$, then $ D_{\perp}(t)\leq D(\hat{\rho}(t),\hat{\sigma}_0)\leq 1$. Therefore an additional time
\begin{equation}
     T_{\rm add}
    =
    \frac{1}{\Delta}
    \log\left(\frac{D_{\perp}(t)}{\eta}\right)
    \leq
    \frac{1}{\Delta}
    \log\left(\frac{1}{\eta}\right)
\end{equation}
is sufficient to reduce this whole support-orthogonal contribution below
accuracy $\eta$. For fixed accuracy $\eta$, this additional time has no
explicit dependence on the number of support-orthogonal modes. Its
system-size dependence is controlled only by $\Delta^{-1}$.

Thus the possible $N_{\rm{mode}}$ factor should be viewed as a conservative artifact of bounding the total trace norm by the sum of the norms of all individual bands. In generic situations, different eigenmode contributions do not align coherently in trace norm. In the exceptional situation where many contributions do add without cancellation because they occupy mutually orthogonal supports, their total magnitude is still bounded by the physical trace distance and decays at least with the Liouvillian gap. Consequently, for fixed $\eta$, this mechanism does not change the rapid-mixing requirement $\Delta^{-1} \leq  \mathcal O[\operatorname{poly}(\log L)] $.

We emphasize the limitation of the above argument. It is not a substitute for a fully uniform trace-norm bound on arbitrary non-normal Liouvillians. For highly non-normal generators, one may in principle construct initial states with large overlaps on many spectral projectors, such that the corresponding decaying eigenmodes contributions add nearly coherently in trace norm and make the
triangle-inequality bound close to sharp. This requires a special alignment of phases and singular-vector structures and is therefore a highly fine-tuned worst-case scenario, rather than the behavior expected for generic physical initial states.

If such collective amplification is allowed, then a rigorous mixing-time bound must explicitly control $N_{\rm band}$, the collective spectral projection norm, or the pseudospectral behavior of the Liouvillian. In the main text, we therefore assume that this artificial mechanism is absent. Under this physically natural assumption, the relevant scaling is governed by the Liouvillian gap and the individual decaying eigenmode trace norm factors $C_j$, rather than by the bare number of decaying eigenmodes.

\appendix
\section{Appendix F: Band-Resolved Formulation of the Mixing-Time Bound}
\label{app:decay_rate_bands}

In the main text, we formulated the mixing-time bound in terms of individual
Liouvillian eigenmodes. Here we describe an equivalent coarse-grained
formulation in terms of decay-rate bands. This language is useful when several modes have nearly degenerate decay rates, when the numerical spectrum is clustered, or when one wants a basis-independent description of degenerate spectral subspaces.

We work in the doubled-space representation and assume, for clarity, that $\hat H^D$ is diagonalizable. Its right and left eigenvectors are defined by
\begin{equation}
    \hat H^D|\psi_{\sigma_j}\rangle
    =
    \epsilon_j|\psi_{\sigma_j}\rangle,
    \qquad
    (\hat H^D)^\dagger|\psi_{\tau_j}\rangle
    =
    \epsilon_j^*|\psi_{\tau_j}\rangle,
\end{equation}
with
\begin{equation}
    \epsilon_j=\alpha_j-i\beta_j,
    \qquad
    \beta_j\geq 0,
\end{equation}
and biorthogonal normalization
\begin{equation}
    \langle \psi_{\tau_i}|\psi_{\sigma_j}\rangle
    =
    \Tr(\hat\tau_i^\dagger\hat\sigma_j)
    =
    \delta_{ij}.
\end{equation}
The steady state corresponds to $j=0$, while all decaying modes have
$\beta_j>0$. The Liouvillian gap is
\begin{equation}
    \Delta=\min_{j>0}\beta_j .
\end{equation}

We partition the nonzero spectrum into decay-rate bands $\{\mathcal B_a\}$, defined by
\begin{equation}
    \Gamma_a\leq \beta_j<\Gamma_{a+1},
    \qquad
    j\in \mathcal B_a,
    \qquad
    \Gamma_1=\Delta .
    \label{eq:app_decay_band_def}
\end{equation}
The corresponding doubled-space spectral projector is
\begin{equation}
    \hat P_a^D
    =
    \sum_{j\in \mathcal B_a}
    |\psi_{\sigma_j}\rangle
    \langle \psi_{\tau_j}| .
    \label{eq:app_band_projector_doubled}
\end{equation}
Equivalently, as a superoperator acting on ordinary operators,
\begin{equation}
    \hat P_a(\hat X)
    =
    \operatorname{unvec}\!\left(
    \hat P_a^D|\psi_X^D\rangle
    \right)
    =
    \sum_{j\in\mathcal B_a}
    \hat\sigma_j
    \Tr(\hat\tau_j^\dagger \hat X).
    \label{eq:app_band_projector_operator}
\end{equation}

For a band containing a single eigenmode, the static projector norm
$\|\hat P_a\|_{1\to 1}$ directly gives the trace-norm factor. For a band containing several non-normal modes, however, the relative phases generated inside the band can change the induced trace norm. It is therefore convenient to define the dynamical band factor
\begin{equation}
    \mathcal{C}_a
    =
    \sup_{t\geq 0}
    e^{\Gamma_a t}
    \left\|
    \hat{\mathcal{T}}_t \hat P_a
    \right\|_{1\to 1},
    \label{eq:app_dynamic_band_prefactor}
\end{equation}
where $ \hat{\mathcal{T}}_t(\hat X) = \operatorname{unvec}\!\left(
    e^{-i\hat H^D t}|\psi_X^D\rangle \right)$ is the physical time-evolution superoperator. By construction,
\begin{equation}
    \left\|
    \hat{\mathcal{T}}_t \hat P_a(\hat X)
    \right\|_1
    \leq
    \mathcal{C}_a e^{-\Gamma_a t}\|\hat X\|_1 .
    \label{eq:app_band_evolution_bound}
\end{equation}

This definition gives a basis-independent measure of the worst-case trace-norm amplification associated with the entire decay band. In the
diagonalizable case, it is bounded by the sum of the individual eigenmode factors. Defining
\begin{equation}
    \hat P_j(\hat X)
    =
    \hat\sigma_j\Tr(\hat\tau_j^\dagger \hat X),
    \qquad
    C_j=\|\hat P_j\|_{1\to 1},
\end{equation}
we have
\begin{equation}
    \mathcal{C}_a
    \leq
    \sum_{j\in\mathcal B_a} C_j .
    \label{eq:app_dynamic_band_prefactor_bound_1}
\end{equation}
Furthermore, by the duality between the trace norm and the operator norm,
\begin{equation}
    C_j
    \leq
    \|\hat\sigma_j\|_1\|\hat\tau_j\|_\infty ,
\end{equation}
and therefore
\begin{equation}
    \mathcal{C}_a
    \leq
    \sum_{j\in\mathcal B_a}
    \|\hat\sigma_j\|_1\|\hat\tau_j\|_\infty .
    \label{eq:app_dynamic_band_prefactor_bound_2}
\end{equation}
If all modes in a band share the same eigenvalue, or if the internal evolution within the band does not generate additional trace-norm amplification, then $\mathcal{C}_a$ reduces, up to constants, to the static projector norm $\|\hat P_a\|_{1\to 1}$.

Let $ \hat X_0=\hat\rho_0-\hat\sigma_0$. For every physical initial state,
\begin{equation}
    \|\hat X_0\|_1
    =
    \|\hat\rho_0-\hat\sigma_0\|_1
    \leq 2 .
    \label{eq:app_physical_trace_bound}
\end{equation}
Using Eq.~\eqref{eq:app_band_evolution_bound}, the contribution of band
$\mathcal B_a$ to the evolved trace distance is bounded by
\begin{equation}
    \frac12
    \left\|
    \hat{\mathcal{T}}_t\hat P_a(\hat X_0)
    \right\|_1
    \leq
    \mathcal{C}_a e^{-\Gamma_a t}.
    \label{eq:app_band_trace_distance_bound}
\end{equation}
Summing over all decay bands gives the conservative estimate
\begin{equation}
    D(\hat\hat{\rho}(t),\hat\sigma_0)
    \leq
    \sum_{a=1}^{N_{\rm band}}
    \mathcal{C}_a e^{-\Gamma_a t}.
    \label{eq:app_band_sum_bound}
\end{equation}
Therefore, a sufficient upper bound on the mixing time is
\begin{equation}
    \tau_{\rm mix}(\eta)
    \lesssim
    \max_a
    \frac{1}{\Gamma_a}
    \log\left(
    \frac{N_{\rm band}\mathcal{C}_a}{\eta}
    \right),
    \label{eq:app_band_mixing_time_bound}
\end{equation}
up to constants of order unity. If Jordan blocks are present, the spectral projectors should be replaced by the corresponding Riesz projectors, and the right-hand side acquires additional polynomial factors in $t$. These factors lead only to additional logarithmic corrections in the sufficient mixing-time bound.

The single-eigenmode formulation used in the main text is recovered by taking each band $\mathcal B_a$ to contain only one eigenmode. The band language is nevertheless useful in several situations. First, for exactly degenerate eigenvalues, individual eigenvectors inside the degenerate subspace are not unique, whereas the spectral projector onto the full degenerate subspace is basis independent. Second, for nearly degenerate decay rates, grouping modes
into a decay band gives a cleaner description of their collective contribution to the trace distance. Third, in numerical calculations, clustered non-normal eigenvalues can make individual eigenvectors ill-conditioned, while the projector onto the whole cluster is often more stable. Finally, the band description can exploit cancellations within a decay sector through the norm of the full band evolution, instead of applying a separate triangle inequality to every individual eigenmode.

Thus, the band formulation is a coarse-grained version of the eigenmode criterion. It does not change the physical content: rapid or fast mixing requires control of both the decay rates $\Gamma_a$ and the associated trace-norm factors $\mathcal{C}_a$. The advantage is that the relevant factor is assigned to an invariant decay sector rather than to a particular choice of eigenbasis within that sector.

\section{Appendix G: Liouvillian Skin Effect and the Trace-Norm Factor}
\label{app:skin_effect_prefactor}

In this Appendix, we explain how the mixing-time scaling found in Ref.\cite{Haga_2021} can be understood from the trace-norm factor appearing in our general bound for mixing time. The essential point is that the mixing time is not controlled only by the Liouvillian gap $\Delta$, but also by the trace norm of the spectral projector associated with the slow mode.

Consider a diagonalizable Liouvillian and a simple eigenvalue $\lambda_j=-\beta_j+ i \alpha_j$. The corresponding rank-one spectral projector is
\begin{equation}
    \hat P_j(\hat X)
    =
    \hat\sigma_j
    \Tr\!\left(
        \hat\tau_j^\dagger \hat X
    \right),
\end{equation}
where $\hat\sigma_j$ and $\hat\tau_j$ are right and left eigenoperators,
\begin{equation}
    \hat{\mathcal{L}}(\hat\sigma_j)=\lambda_j\hat\sigma_j,
    \qquad
    \hat{\mathcal{L}}^\dagger(\hat\tau_j)=\lambda_j^*\hat\tau_j,
\end{equation}
normalized bi-orthogonally as $\Tr\!\left(
        \hat\tau_i^\dagger \hat\sigma_j
    \right)
    =
    \delta_{ij}$. 
    
The trace-norm factor of this mode is
\begin{equation}
    C_j = \|\hat P_j\|_{1\to 1}.
\end{equation}
Using trace-norm duality, one obtains
\begin{equation}
    C_j =  \|\hat\sigma_j\|_1  \|\hat\tau_j\|_\infty .
    \label{eq:skin_Cj_def}
\end{equation}
Thus $C_j$ measures the possible non-normal amplification of the Liouvillian eigenmode.

We now apply this observation to systems with the Liouvillian skin effect. We use the convention
\begin{equation}
    \Tr\!\left(
        \hat\sigma_j^\dagger \hat\sigma_j
    \right)
    =
    1 .
\end{equation}
Let $\hat{\bar\tau}_j$ be a naturally normalized left eigenoperator, for example satisfying
\begin{equation}
    \Tr\!\left(
        \hat{\bar\tau}_j^\dagger \hat{\bar\tau}_j
    \right)
    =
    1 .
\end{equation}
In the presence of the Liouvillian skin effect, the right and left eigenoperators can be localized near opposite boundaries. Their overlap can therefore be exponentially small in the system size,
\begin{equation}
    s_j
    \equiv
    \Tr\!\left(
        \hat{\bar\tau}_j^\dagger \hat\sigma_j
    \right)
    \sim
    e^{-aL/\xi},
\end{equation}
where $\xi$ is the localization length and $a$ is a nonuniversal constant of order one.

However, the left eigenoperator entering the spectral projector is the biorthogonally normalized one. It must satisfy $\Tr\!\left(
        \hat\tau_j^\dagger \hat\sigma_j \right)  =  1$. Therefore, up to an irrelevant phase convention,
\begin{equation}
    \hat\tau_j  =   \frac{\hat{\bar\tau}_j}{s_j^*}.
\end{equation}
Its operator norm is enhanced by the inverse overlap,
\begin{equation}
    \|\hat\tau_j\|_\infty  =  \frac{     \|\hat{\bar\tau}_j\|_\infty  }{ |s_j| }  \sim  e^{aL/\xi}  \|\hat{\bar\tau}_j\|_\infty .
\end{equation}
Substituting this into Eq.~\eqref{eq:skin_Cj_def} gives
\begin{equation}
    C_j =
    \|\hat\sigma_j\|_1
    \|\hat\tau_j\|_\infty
    \sim
    e^{aL/\xi}
    \|\hat\sigma_j\|_1
    \|\hat{\bar\tau}_j\|_\infty .
\end{equation}
Therefore, provided the individual norms $\|\hat\sigma_j\|_1$ and
$\|\hat{\bar\tau}_j\|_\infty$ do not compensate the exponentially small overlap, the spectral-projector prefactor scales as
\begin{equation}
    C_j  \sim e^{O(L/\xi)} .
\end{equation}
This enhancement is not necessarily due to an exponentially large trace norm of the right eigenoperator alone. Rather, it is a bi-orthogonal effect: the exponentially small overlap between oppositely localized left and right skin modes forces the bi-orthogonally normalized left eigenoperator to have an exponentially large norm.

For the slowest decaying mode, $\beta_1=\Delta$, the single-mode form of our mixing-time bound gives
\begin{equation}
    \tau_{\rm mix}(\eta)
    \lesssim
    \frac{1}{\Delta}
    \log\left(
        \frac{C_1}{\eta}
    \right).
\end{equation}
If $C_1\sim e^{aL/\xi}$, then for fixed accuracy $\eta$,
\begin{equation}
    \tau_{\rm mix}(\eta)
    \sim
    \frac{1}{\Delta}
    \left(
        \log\frac{1}{\eta}
        +
        \frac{aL}{\xi}
    \right)
    \sim
    \Delta^{-1}
    \left(
        1+\frac{L}{\xi}
    \right),
\end{equation}
up to nonuniversal constants and convention-dependent definitions of the localization length. This reproduces the scaling reported in Ref.\cite{Haga_2021}.

We emphasize that this argument describes a particular mechanism for slow mixing: the dominant slow mode has an exponentially large spectral-projector prefactor because of the bi-orthogonal non-orthogonality of left and right eigenmodes. In more general open systems, other decaying modes or spectral bands may also contribute. The general mixing-time bound therefore requires the corresponding prefactors $C_j$, or their band-resolved analogues $\mathcal{C}_a$, to be controlled for all relevant decaying modes.

\bibliography{ref.bib}

@article{Vermersch_2019,
   title={Probing Scrambling Using Statistical Correlations between Randomized Measurements},
   volume={9},
   ISSN={2160-3308},
   url={http://dx.doi.org/10.1103/PhysRevX.9.021061},
   DOI={10.1103/physrevx.9.021061},
   number={2},
   journal={Physical Review X},
   publisher={American Physical Society (APS)},
   author={Vermersch, B. and Elben, A. and Sieberer, L.M. and Yao, N.Y. and Zoller, P.},
   year={2019},
   month=jun }

@article{Maldacena:2015waa, 
    title={A bound on chaos}, 
    volume={2016}, 
    ISSN={1029-8479}, 
    url={http://dx.doi.org/10.1007/JHEP08(2016)106}, 
    DOI={10.1007/jhep08(2016)106}, 
    number={8}, 
    journal={Journal of High Energy Physics}, publisher={Springer Science and Business Media LLC}, 
    author={Maldacena, Juan and Shenker, Stephen H. and Stanford, Douglas}, year={2016}, 
    month=aug 
}

@book{Nielsen_Chuang_2010, place={Cambridge}, title={Quantum Computation and Quantum Information: 10th Anniversary Edition}, publisher={Cambridge University Press}, author={Nielsen, Michael A. and Chuang, Isaac L.}, year={2010},
url={},
DOI={} }

@article{PhysRevResearch.3.043060,
  title = {R\'enyi entropy dynamics and Lindblad spectrum for open quantum systems},
  author = {Zhou, Yi-Neng and Mao, Liang and Zhai, Hui},
  journal = {Phys. Rev. Res.},
  volume = {3},
  issue = {4},
  pages = {043060},
  numpages = {6},
  year = {2021},
  month = {Oct},
  publisher = {American Physical Society},
  doi = {10.1103/PhysRevResearch.3.043060},
  url = {https://link.aps.org/doi/10.1103/PhysRevResearch.3.043060}
}

@misc{lin2025dissipative,
      title={Dissipative Preparation of Many-Body Quantum States: Towards Practical Quantum Advantage}, 
      author={Lin Lin},
      year={2025},
      eprint={2505.21308},
      archivePrefix={arXiv},
      primaryClass={quant-ph},
      url={https://arxiv.org/abs/2505.21308}, 
}

@article{Ding_2024,
   title={Single-ancilla ground state preparation via Lindbladians},
   volume={6},
   ISSN={2643-1564},
   url={http://dx.doi.org/10.1103/PhysRevResearch.6.033147},
   DOI={10.1103/physrevresearch.6.033147},
   number={3},
   journal={Physical Review Research},
   publisher={American Physical Society (APS)},
   author={Ding, Zhiyan and Chen, Chi-Fang and Lin, Lin},
   year={2024},
   month=aug }

@article{zhan2025rapidquantumgroundstate,
   title={Rapid Quantum Ground State Preparation via Dissipative Dynamics},
   volume={16},
   ISSN={2160-3308},
   url={http://dx.doi.org/10.1103/wzb3-dbg9},
   DOI={10.1103/wzb3-dbg9},
   number={1},
   journal={Physical Review X},
   publisher={American Physical Society (APS)},
   author={Zhan, Yongtao and Ding, Zhiyan and Huhn, Jakob and Gray, Johnnie and Preskill, John and Chan, Garnet Kin-Lic and Lin, Lin},
   year={2026},
   month=Jan }

@article{Zhou_2024,
   title={General properties of the spectral form factor in open quantum systems},
   volume={19},
   ISSN={2095-0470},
   url={http://dx.doi.org/10.1007/s11467-024-1406-7},
   DOI={10.1007/s11467-024-1406-7},
   number={3},
   journal={Frontiers of Physics},
   publisher={China Engineering Science Press Co. Ltd.},
   author={Zhou, Yi-Neng and Zhou, Tian-Gang and Zhang, Pengfei},
   year={2024},
   month=apr }

@article{TysonOperatorSchmidtDecompositionsFourier2003,
  title = {Operator-{{Schmidt}} Decompositions and the {{Fourier}} Transform, with Applications to the Operator-{{Schmidt}} Numbers of Unitaries},
  author = {Tyson, Jon E.},
  year = {2003},
  month = sep,
  journal = {J. Phys. A: Math. Gen.},
  volume = {36},
  number = {39},
  pages = {10101},
  issn = {0305-4470},
  doi = {10.1088/0305-4470/36/39/309},
  urldate = {2023-03-25}
}

@article{VidalMixedStateDynamicsOneDimensional2004,
  title = {Mixed-{{State Dynamics}} in {{One-Dimensional Quantum Lattice Systems}}: {{A Time-Dependent Superoperator Renormalization Algorithm}}},
  shorttitle = {Mixed-{{State Dynamics}} in {{One-Dimensional Quantum Lattice Systems}}},
  author = {Zwolak, Michael and Vidal, Guifr{\'e}},
  year = {2004},
  month = nov,
  journal = {Phys. Rev. Lett.},
  volume = {93},
  number = {20},
  pages = {207205},
  publisher = {{American Physical Society}},
  doi = {10.1103/PhysRevLett.93.207205},
  urldate = {2023-03-25}
}

@article{Fang_2025,
   title={Mixing Time of Open Quantum Systems via Hypocoercivity},
   volume={134},
   ISSN={1079-7114},
   url={http://dx.doi.org/10.1103/PhysRevLett.134.140405},
   DOI={10.1103/physrevlett.134.140405},
   number={14},
   journal={Physical Review Letters},
   publisher={American Physical Society (APS)},
   author={Fang, Di and Lu, Jianfeng and Tong, Yu},
   year={2025},
   month=apr }

@article{Haga_2021,
   title={Liouvillian Skin Effect: Slowing Down of Relaxation Processes without Gap Closing},
   volume={127},
   ISSN={1079-7114},
   url={http://dx.doi.org/10.1103/PhysRevLett.127.070402},
   DOI={10.1103/physrevlett.127.070402},
   number={7},
   journal={Physical Review Letters},
   publisher={American Physical Society (APS)},
   author={Haga, Taiki and Nakagawa, Masaya and Hamazaki, Ryusuke and Ueda, Masahito},
   year={2021},
   month=aug }

@article{Kastoryano_2013,
   title={Quantum logarithmic Sobolev inequalities and rapid mixing},
   volume={54},
   ISSN={1089-7658},
   url={http://dx.doi.org/10.1063/1.4804995},
   DOI={10.1063/1.4804995},
   number={5},
   journal={Journal of Mathematical Physics},
   publisher={AIP Publishing},
   author={Kastoryano, Michael J. and Temme, Kristan},
   year={2013},
   month=may }

@misc{capel2021modifiedlogarithmicsobolevinequality,
  author       = {Capel, {\'A}ngela and Rouz{\'e}, Cambyse and Stilck Fran{\c{c}}a, Daniel},
  title        = {The modified logarithmic {S}obolev inequality for quantum spin systems: classical and commuting nearest neighbour interactions},
  year         = {2021},
  eprint       = {2009.11817},
  archivePrefix= {arXiv},
  primaryClass = {quant-ph},
  url          = {https://arxiv.org/abs/2009.11817}
}

@article{Bardet_2023,
  author    = {Bardet, Ivan and Capel, {\'A}ngela and Gao, Li and Lucia, Angelo
               and P{\'e}rez-Garc{\'\i}a, David and Rouz{\'e}, Cambyse},
  title     = {Rapid Thermalization of Spin Chain Commuting Hamiltonians},
  journal   = {Physical Review Letters},
  volume    = {130},
  number    = {6},
  pages     = {060401},
  year      = {2023},
  month     = feb,
  publisher = {American Physical Society},
  doi       = {10.1103/PhysRevLett.130.060401},
  url       = {https://doi.org/10.1103/PhysRevLett.130.060401}
}

@article{Bardet_2021,
  author    = {Bardet, Ivan and Capel, {\'A}ngela and Lucia, Angelo
               and P{\'e}rez-Garc{\'\i}a, David and Rouz{\'e}, Cambyse},
  title     = {On the modified logarithmic Sobolev inequality for the heat-bath dynamics for 1D systems},
  journal   = {Journal of Mathematical Physics},
  year      = {2021},
  volume    = {62},
  number    = {6},
  pages     = {061901},
  month     = jun,
  publisher = {AIP Publishing},
  doi       = {10.1063/1.5142186},
  url       = {https://doi.org/10.1063/1.5142186}
}

@article{Gao:2021xaw,
    author = "Gao, Li and Rouz{\'e}, Cambyse",
    title = "{Complete Entropic Inequalities for Quantum Markov Chains}",
    eprint = "2102.04146",
    archivePrefix = "arXiv",
    primaryClass = "quant-ph",
    doi = "10.1007/s00205-022-01785-1",
    journal = "Arch. Ration. Mech. Anal.",
    volume = "245",
    number = "1",
    pages = "183--238",
    year = "2022"
}

@article{PhysRevResearch.6.033147,
  title = {Single-ancilla ground state preparation via Lindbladians},
  author = {Ding, Zhiyan and Chen, Chi-Fang and Lin, Lin},
  journal = {Phys. Rev. Res.},
  volume = {6},
  issue = {3},
  pages = {033147},
  numpages = {24},
  year = {2024},
  month = {Aug},
  publisher = {American Physical Society},
  doi = {10.1103/PhysRevResearch.6.033147},
  url = {https://link.aps.org/doi/10.1103/PhysRevResearch.6.033147}
}

@article{PhysRevA.77.042312,
  title = {Mixing times in quantum walks on the hypercube},
  author = {Marquezino, F. L. and Portugal, R. and Abal, G. and Donangelo, R.},
  journal = {Phys. Rev. A},
  volume = {77},
  issue = {4},
  pages = {042312},
  numpages = {8},
  year = {2008},
  month = {Apr},
  publisher = {American Physical Society},
  doi = {10.1103/PhysRevA.77.042312},
  url = {https://link.aps.org/doi/10.1103/PhysRevA.77.042312}
}

@article{PhysRevE.92.042143,
  title = {Relaxation times of dissipative many-body quantum systems},
  author = {\ifmmode \check{Z}\else \v{Z}\fi{}nidari\ifmmode \check{c}\else \v{c}\fi{}, Marko},
  journal = {Phys. Rev. E},
  volume = {92},
  issue = {4},
  pages = {042143},
  numpages = {17},
  year = {2015},
  month = {Oct},
  publisher = {American Physical Society},
  doi = {10.1103/PhysRevE.92.042143},
  url = {https://link.aps.org/doi/10.1103/PhysRevE.92.042143}
}

@article{Syzranov_2018,
   title={Out-of-time-order correlators in finite open systems},
   volume={97},
   ISSN={2469-9969},
   url={http://dx.doi.org/10.1103/PhysRevB.97.161114},
   DOI={10.1103/physrevb.97.161114},
   number={16},
   journal={Physical Review B},
   publisher={American Physical Society (APS)},
   author={Syzranov, S. V. and Gorshkov, A. V. and Galitski, V.},
   year={2018},
   month=apr }

@article{Han_2022,
   title={Quantum Information Scrambling in Non-Markovian Open Quantum System},
   volume={24},
   ISSN={1099-4300},
   url={http://dx.doi.org/10.3390/e24111532},
   DOI={10.3390/e24111532},
   number={11},
   journal={Entropy},
   publisher={MDPI AG},
   author={Han, Li-Ping and Zou, Jian and Li, Hai and Shao, Bin},
   year={2022},
   month=oct, pages={1532} }

@misc{fang2024probingcriticalphenomenaopen,
      title={Probing critical phenomena in open quantum systems using atom arrays}, 
      author={Fang Fang and Kenneth Wang and Vincent S. Liu and Yu Wang and Ryan Cimmino and Julia Wei and Marcus Bintz and Avery Parr and Jack Kemp and Kang-Kuen Ni and Norman Y. Yao},
      year={2024},
      eprint={2402.15376},
      archivePrefix={arXiv},
      primaryClass={quant-ph},
      url={https://arxiv.org/abs/2402.15376}, 
}

@article{PhysRevLett.131.160402,
  title = {Operator Growth in Open Quantum Systems},
  author = {Schuster, Thomas and Yao, Norman Y.},
  journal = {Phys. Rev. Lett.},
  volume = {131},
  issue = {16},
  pages = {160402},
  numpages = {6},
  year = {2023},
  month = {Oct},
  publisher = {American Physical Society},
  doi = {10.1103/PhysRevLett.131.160402},
  url = {https://link.aps.org/doi/10.1103/PhysRevLett.131.160402}
}

@article{Bhattacharya_2022,
   title={Operator growth and Krylov construction in dissipative open quantum systems},
   volume={2022},
   ISSN={1029-8479},
   url={http://dx.doi.org/10.1007/JHEP12(2022)081},
   DOI={10.1007/jhep12(2022)081},
   number={12},
   journal={Journal of High Energy Physics},
   publisher={Springer Science and Business Media LLC},
   author={Bhattacharya, Aranya and Nandy, Pratik and Nath, Pingal Pratyush and Sahu, Himanshu},
   year={2022},
   month=dec }

@article{PhysRevResearch.5.033085,
  title = {Krylov complexity in open quantum systems},
  author = {Liu, Chang and Tang, Haifeng and Zhai, Hui},
  journal = {Phys. Rev. Res.},
  volume = {5},
  issue = {3},
  pages = {033085},
  numpages = {7},
  year = {2023},
  month = {Aug},
  publisher = {American Physical Society},
  doi = {10.1103/PhysRevResearch.5.033085},
  url = {https://link.aps.org/doi/10.1103/PhysRevResearch.5.033085}
}

@book{Breuer2002,
  author    = {Heinz-Peter Breuer and Francesco Petruccione},
  title     = {The Theory of Open Quantum Systems},
  publisher = {Oxford University Press},
  address   = {Oxford},
  year      = {2002},
  isbn      = {978-0-19-852063-4},      
}

@book{Gardiner2004,
  author    = {C. W. Gardiner and Peter Zoller},
  title     = {Quantum Noise: A Handbook of Markovian and Non-Markovian Quantum Stochastic Methods with Applications to Quantum Optics},
  edition   = {3},
  year      = {2004},
  series    = {Springer Series in Synergetics},
  publisher = {Springer},
  address   = {Berlin, Heidelberg},
  isbn      = {978-3-540-22301-6}
}

@article{lindblad1976generators,
  title={On the generators of quantum dynamical semigroups},
  author={Lindblad, Goran},
  journal={Communications in mathematical physics},
  volume={48},
  number={2},
  pages={119--130},
  year={1976},
  publisher={Springer}
}

@article{Gorini:1975nb,
    author = "Gorini, Vittorio and Kossakowski, Andrzej and Sudarshan, E. C. G.",
    title = "{Completely Positive Dynamical Semigroups of N Level Systems}",
    reportNumber = "CPT-244-TEXAS, ORO-3992-200",
    doi = "10.1063/1.522979",
    journal = "J. Math. Phys.",
    volume = "17",
    pages = "821",
    year = "1976"
}

@article{Mi_2021,
   title={Information scrambling in quantum circuits},
   volume={374},
   ISSN={1095-9203},
   url={http://dx.doi.org/10.1126/science.abg5029},
   DOI={10.1126/science.abg5029},
   number={6574},
   journal={Science},
   publisher={American Association for the Advancement of Science (AAAS)},
   author={Mi, Xiao and Roushan, Pedram and Quintana, Chris and Mandrà, Salvatore and Marshall, Jeffrey and Neill, Charles and Arute, Frank and Arya, Kunal and Atalaya, Juan and Babbush, Ryan and Bardin, Joseph C. and Barends, Rami and Basso, Joao and Bengtsson, Andreas and Boixo, Sergio and Bourassa, Alexandre and Broughton, Michael and Buckley, Bob B. and Buell, David A. and Burkett, Brian and Bushnell, Nicholas and Chen, Zijun and Chiaro, Benjamin and Collins, Roberto and Courtney, William and Demura, Sean and Derk, Alan R. and Dunsworth, Andrew and Eppens, Daniel and Erickson, Catherine and Farhi, Edward and Fowler, Austin G. and Foxen, Brooks and Gidney, Craig and Giustina, Marissa and Gross, Jonathan A. and Harrigan, Matthew P. and Harrington, Sean D. and Hilton, Jeremy and Ho, Alan and Hong, Sabrina and Huang, Trent and Huggins, William J. and Ioffe, L. B. and Isakov, Sergei V. and Jeffrey, Evan and Jiang, Zhang and Jones, Cody and Kafri, Dvir and Kelly, Julian and Kim, Seon and Kitaev, Alexei and Klimov, Paul V. and Korotkov, Alexander N. and Kostritsa, Fedor and Landhuis, David and Laptev, Pavel and Lucero, Erik and Martin, Orion and McClean, Jarrod R. and McCourt, Trevor and McEwen, Matt and Megrant, Anthony and Miao, Kevin C. and Mohseni, Masoud and Montazeri, Shirin and Mruczkiewicz, Wojciech and Mutus, Josh and Naaman, Ofer and Neeley, Matthew and Newman, Michael and Niu, Murphy Yuezhen and O’Brien, Thomas E. and Opremcak, Alex and Ostby, Eric and Pato, Balint and Petukhov, Andre and Redd, Nicholas and Rubin, Nicholas C. and Sank, Daniel and Satzinger, Kevin J. and Shvarts, Vladimir and Strain, Doug and Szalay, Marco and Trevithick, Matthew D. and Villalonga, Benjamin and White, Theodore and Yao, Z. Jamie and Yeh, Ping and Zalcman, Adam and Neven, Hartmut and Aleiner, Igor and Kechedzhi, Kostyantyn and Smelyanskiy, Vadim and Chen, Yu},
   year={2021},
   month=dec, pages={1479–1483} }

@article{Landsman_2019,
   title={Verified quantum information scrambling},
   volume={567},
   ISSN={1476-4687},
   url={http://dx.doi.org/10.1038/s41586-019-0952-6},
   DOI={10.1038/s41586-019-0952-6},
   number={7746},
   journal={Nature},
   publisher={Springer Science and Business Media LLC},
   author={Landsman, K. A. and Figgatt, C. and Schuster, T. and Linke, N. M. and Yoshida, B. and Yao, N. Y. and Monroe, C.},
   year={2019},
   month=mar, pages={61–65} }

@article{48651,title	= {Quantum Supremacy using a Programmable Superconducting Processor},author	= {Frank Arute and Kunal Arya and Ryan Babbush and Dave Bacon and Joseph Bardin and Rami Barends and Rupak Biswas and Sergio Boixo and Fernando Brandao and David Buell and Brian Burkett and Yu Chen and Jimmy Chen and Ben Chiaro and Roberto Collins and William Courtney and Andrew Dunsworth and Edward Farhi and Brooks Foxen and Austin Fowler and Craig Michael Gidney and Marissa Giustina and Rob Graff and Keith Guerin and Steve Habegger and Matthew Harrigan and Michael Hartmann and Alan Ho and Markus Rudolf Hoffmann and Trent Huang and Travis Humble and Sergei Isakov and Evan Jeffrey and Zhang Jiang and Dvir Kafri and Kostyantyn Kechedzhi and Julian Kelly and Paul Klimov and Sergey Knysh and Alexander Korotkov and Fedor Kostritsa and Dave Landhuis and Mike Lindmark and Erik Lucero and Dmitry Lyakh and Salvatore Mandrà and Jarrod Ryan McClean and Matthew McEwen and Anthony Megrant and Xiao Mi and Kristel Michielsen and Masoud Mohseni and Josh Mutus and Ofer Naaman and Matthew Neeley and Charles Neill and Murphy Yuezhen Niu and Eric Ostby and Andre Petukhov and John Platt and Chris Quintana and Eleanor G. Rieffel and Pedram Roushan and Nicholas Rubin and Daniel Sank and Kevin J. Satzinger and Vadim Smelyanskiy and Kevin Jeffery Sung and Matt Trevithick and Amit Vainsencher and Benjamin Villalonga and Ted White and Z. Jamie Yao and Ping Yeh and Adam Zalcman and Hartmut Neven and John Martinis},year	= {2019},URL	= {https://www.nature.com/articles/s41586-019-1666-5},journal	= {Nature},pages	= {505–510},volume	= {574}}

@article{Shenker_2014,
   title={Black holes and the butterfly effect},
   volume={2014},
   ISSN={1029-8479},
   url={http://dx.doi.org/10.1007/JHEP03(2014)067},
   DOI={10.1007/jhep03(2014)067},
   number={3},
   journal={Journal of High Energy Physics},
   publisher={Springer Science and Business Media LLC},
   author={Shenker, Stephen H. and Stanford, Douglas},
   year={2014},
   month=mar }

@ARTICLE{1969JETP...28.1200L,
       author = {{Larkin}, A.~I. and {Ovchinnikov}, Yu. N.},
        title = "{Quasiclassical Method in the Theory of Superconductivity}",
      journal = {Soviet Journal of Experimental and Theoretical Physics},
         year = 1969,
        month = jun,
       volume = {28},
        pages = {1200},
       adsurl = {https://ui.adsabs.harvard.edu/abs/1969JETP...28.1200L}
}

@ARTICLE{2018NatPh..14..988S,
       author = {{Swingle}, Brian},
        title = "{Unscrambling the physics of out-of-time-order correlators}",
      journal = {Nature Physics},
         year = 2018,
        month = oct,
       volume = {14},
       number = {10},
        pages = {988-990},
          doi = {10.1038/s41567-018-0295-5},
       adsurl = {https://ui.adsabs.harvard.edu/abs/2018NatPh..14..988S}
}

@article{RevModPhys.59.1,
  title = {Dynamics of the dissipative two-state system},
  author = {Leggett, A. J. and Chakravarty, S. and Dorsey, A. T. and Fisher, Matthew P. A. and Garg, Anupam and Zwerger, W.},
  journal = {Rev. Mod. Phys.},
  volume = {59},
  issue = {1},
  pages = {1--85},
  numpages = {0},
  year = {1987},
  month = {Jan},
  publisher = {American Physical Society},
  doi = {10.1103/RevModPhys.59.1},
  url = {https://link.aps.org/doi/10.1103/RevModPhys.59.1}
}

@article{Raftery_2014,
   title={Observation of a Dissipation-Induced Classical to Quantum Transition},
   volume={4},
   ISSN={2160-3308},
   url={http://dx.doi.org/10.1103/PhysRevX.4.031043},
   DOI={10.1103/physrevx.4.031043},
   number={3},
   journal={Physical Review X},
   publisher={American Physical Society (APS)},
   author={Raftery, J. and Sadri, D. and Schmidt, S. and Türeci, H.E. and Houck, A.A.},
   year={2014},
   month=sep }

@article{Carusotto_2013,
   title={Quantum fluids of light},
   volume={85},
   ISSN={1539-0756},
   url={http://dx.doi.org/10.1103/RevModPhys.85.299},
   DOI={10.1103/revmodphys.85.299},
   number={1},
   journal={Reviews of Modern Physics},
   publisher={American Physical Society (APS)},
   author={Carusotto, Iacopo and Ciuti, Cristiano},
   year={2013},
   month=feb, pages={299–366} }

@article{Schmitt_2020,
   title={Quantum Many-Body Dynamics in Two Dimensions with Artificial Neural Networks},
   volume={125},
   ISSN={1079-7114},
   url={http://dx.doi.org/10.1103/PhysRevLett.125.100503},
   DOI={10.1103/physrevlett.125.100503},
   number={10},
   journal={Physical Review Letters},
   publisher={American Physical Society (APS)},
   author={Schmitt, Markus and Heyl, Markus},
   year={2020},
   month=sep }

@article{PhysRevLett.77.4728,
  title = {Quantum Reservoir Engineering with Laser Cooled Trapped Ions},
  author = {Poyatos, J. F. and Cirac, J. I. and Zoller, P.},
  journal = {Phys. Rev. Lett.},
  volume = {77},
  issue = {23},
  pages = {4728--4731},
  numpages = {0},
  year = {1996},
  month = {Dec},
  publisher = {American Physical Society},
  doi = {10.1103/PhysRevLett.77.4728},
  url = {https://link.aps.org/doi/10.1103/PhysRevLett.77.4728}
}

@article{Diehl_2008,
   title={Quantum states and phases in driven open quantum systems with cold atoms},
   volume={4},
   ISSN={1745-2481},
   url={http://dx.doi.org/10.1038/nphys1073},
   DOI={10.1038/nphys1073},
   number={11},
   journal={Nature Physics},
   publisher={Springer Science and Business Media LLC},
   author={Diehl, S. and Micheli, A. and Kantian, A. and Kraus, B. and Büchler, H. P. and Zoller, P.},
   year={2008},
   month=sep, pages={878–883} }

@article{NP2009,
author = {Verstraete, Frank and Wolf, Michael and Cirac, J.},
year = {2009},
month = {09},
pages = {633-636},
title = {Quantum computation and quantum-state engineering driven by dissipation},
volume = {5},
journal = {Nature Physics},
doi = {10.1038/nphys1342}
}

@article{Kraus_2008,
   title={Preparation of entangled states by quantum Markov processes},
   volume={78},
   ISSN={1094-1622},
   url={http://dx.doi.org/10.1103/PhysRevA.78.042307},
   DOI={10.1103/physreva.78.042307},
   number={4},
   journal={Physical Review A},
   publisher={American Physical Society (APS)},
   author={Kraus, B. and Büchler, H. P. and Diehl, S. and Kantian, A. and Micheli, A. and Zoller, P.},
   year={2008},
   month=oct }

@article{PhysRevLett.107.080503,
  title = {Entanglement Generated by Dissipation and Steady State Entanglement of Two Macroscopic Objects},
  author = {Krauter, Hanna and Muschik, Christine A. and Jensen, Kasper and Wasilewski, Wojciech and Petersen, Jonas M. and Cirac, J. Ignacio and Polzik, Eugene S.},
  journal = {Phys. Rev. Lett.},
  volume = {107},
  issue = {8},
  pages = {080503},
  numpages = {5},
  year = {2011},
  month = {Aug},
  publisher = {American Physical Society},
  doi = {10.1103/PhysRevLett.107.080503},
  url = {https://link.aps.org/doi/10.1103/PhysRevLett.107.080503}
}

@article{f409a97a-1b7e-38fc-a741-afe96ecf8ad1,
 ISSN = {00368075, 10959203},
 URL = {http://www.jstor.org/stable/2899535},
 author = {Seth Lloyd},
 journal = {Science},
 number = {5278},
 pages = {1073--1078},
 publisher = {American Association for the Advancement of Science},
 title = {Universal Quantum Simulators},
 urldate = {2025-08-23},
 volume = {273},
 year = {1996}
}

@article{RevModPhys.86.153,
  title = {Quantum simulation},
  author = {Georgescu, I. M. and Ashhab, S. and Nori, Franco},
  journal = {Rev. Mod. Phys.},
  volume = {86},
  issue = {1},
  pages = {153--185},
  numpages = {33},
  year = {2014},
  month = {Mar},
  publisher = {American Physical Society},
  doi = {10.1103/RevModPhys.86.153},
  url = {https://link.aps.org/doi/10.1103/RevModPhys.86.153}
}

@ARTICLE{2012NatPh...8..267B,
       author = {{Bloch}, Immanuel and {Dalibard}, Jean and {Nascimb{\`e}ne}, Sylvain},
        title = "{Quantum simulations with ultracold quantum gases}",
      journal = {Nature Physics},
         year = 2012,
        month = apr,
       volume = {8},
       number = {4},
        pages = {267-276},
          doi = {10.1038/nphys2259},
       adsurl = {https://ui.adsabs.harvard.edu/abs/2012NatPh...8..267B}
}

@article{Greiner2002,
author = {Greiner, Markus and Mandel, Olaf and Esslinger, Tilman and Haensch, Theodor and Bloch, Immanuel},
year = {2002},
month = {02},
pages = {39-44},
title = {Quantum phase transition from a superfluid to a Mott insulator in a gas of ultracold atoms},
volume = {415},
journal = {Nature},
doi = {10.1038/415039a}
}

@article{Lanyon_2011,
   title={Universal Digital Quantum Simulation with Trapped Ions},
   volume={334},
   ISSN={1095-9203},
   url={http://dx.doi.org/10.1126/science.1208001},
   DOI={10.1126/science.1208001},
   number={6052},
   journal={Science},
   publisher={American Association for the Advancement of Science (AAAS)},
   author={Lanyon, B. P. and Hempel, C. and Nigg, D. and Müller, M. and Gerritsma, R. and Zähringer, F. and Schindler, P. and Barreiro, J. T. and Rambach, M. and Kirchmair, G. and Hennrich, M. and Zoller, P. and Blatt, R. and Roos, C. F.},
   year={2011},
   month=oct, pages={57–61} }

@article{Martinez_2016,
   title={Real-time dynamics of lattice gauge theories with a few-qubit quantum computer},
   volume={534},
   ISSN={1476-4687},
   url={http://dx.doi.org/10.1038/nature18318},
   DOI={10.1038/nature18318},
   number={7608},
   journal={Nature},
   publisher={Springer Science and Business Media LLC},
   author={Martinez, Esteban A. and Muschik, Christine A. and Schindler, Philipp and Nigg, Daniel and Erhard, Alexander and Heyl, Markus and Hauke, Philipp and Dalmonte, Marcello and Monz, Thomas and Zoller, Peter and Blatt, Rainer},
   year={2016},
   month=jun, pages={516–519} }

@article{Feynman1982,
author = {Feynman, Richard P.},
title = {Simulating physics with computers},
journal = {International Journal of Theoretical Physics},
volume = {21},
number = {6-7},
pages = {467--488},
year = {1982},
publisher = {Springer US}
}

@inproceedings{Grover1996,
  author    = {Lov K. Grover},
  title     = {A fast quantum mechanical algorithm for database search},
  booktitle = {Proceedings of the twenty-eighth annual {ACM} symposium on Theory of computing},
  pages     = {212--219},
  year      = {1996},
  doi       = {10.1145/237814.237866},
  publisher = {{ACM}},
  address   = {Philadelphia, Pennsylvania, USA}
}

@inproceedings{Aharonov2001,
  author    = {Dorit Aharonov and
               Andris Ambainis and
               Julia Kempe and
               Umesh Vazirani},
  title     = {Quantum Walks on Graphs},
  booktitle = {Proceedings of the Thirty-Third Annual {ACM} Symposium on Theory of
               Computing, {STOC} 2001, Hersonissos, Crete, Greece, July 6-8, 2001},
  pages     = {50--59},
  year      = {2001},
  doi       = {10.1145/380752.380758},
  publisher = {{ACM}},
  url       = {https://doi.org/10.1145/380752.380758}
}

@book{Levin2009,
  author    = {Levin, David A. and
               Peres, Yuval and
               Wilmer, Elizabeth L.},
  title     = {Markov Chains and Mixing Times},
  publisher = {American Mathematical Society},
  year      = {2009},
  series    = {Fields Institute Monographs},
  volume    = {36},
  isbn      = {978-0-8218-4739-8}
}

@article{Diaconis1993,
  author  = {Diaconis, Persi and Saloff-Coste, Laurent},
  title   = {A Note on the Convergence of Markov Chains},
  journal = {The Annals of Applied Probability},
  volume  = {3},
  number  = {3},
  pages   = {696--708},
  year    = {1993},
  doi     = {10.1214/aoap/1177005436}
}

@book{Huang1987,
  author    = {Huang, Kerson},
  title     = {Statistical Mechanics},
  edition   = {2nd},
  publisher = {John Wiley \& Sons},
  year      = {1987},
  isbn      = {978-0471815181}
}

@article{Lesanovsky2013,
  author  = {Lesanovsky, Igor and B{\"u}chler, Hans Peter},
  title   = {Dissipative phase transitions and non-equilibrium steady states of a spin chain},
  journal = {Journal of Statistical Physics},
  volume  = {151},
  number  = {3-4},
  pages   = {523--546},
  year    = {2013},
  doi     = {10.1007/s10955-013-0740-4}
}

@article{PhysRevA.90.052109,
  title = {Dissipative phase transitions: Independent versus collective decay and spin squeezing},
  author = {Lee, Tony E. and Chan, Ching-Kit and Yelin, Susanne F.},
  journal = {Phys. Rev. A},
  volume = {90},
  issue = {5},
  pages = {052109},
  numpages = {8},
  year = {2014},
  month = {Nov},
  publisher = {American Physical Society},
  doi = {10.1103/PhysRevA.90.052109},
  url = {https://link.aps.org/doi/10.1103/PhysRevA.90.052109}
}

@article{PhysRevA.98.042118,
  title = {Spectral theory of Liouvillians for dissipative phase transitions},
  author = {Minganti, Fabrizio and Biella, Alberto and Bartolo, Nicola and Ciuti, Cristiano},
  journal = {Phys. Rev. A},
  volume = {98},
  issue = {4},
  pages = {042118},
  numpages = {13},
  year = {2018},
  month = {Oct},
  publisher = {American Physical Society},
  doi = {10.1103/PhysRevA.98.042118},
  url = {https://link.aps.org/doi/10.1103/PhysRevA.98.042118}
}

@Article{10.21468/SciPostPhys.20.3.072,
	title={{Generalized Loschmidt echo and information scrambling in open systems}},
	author={Yi-Neng Zhou and Chang Liu},
	journal={SciPost Phys.},
	volume={20},
	pages={072},
	year={2026},
	publisher={SciPost},
	doi={10.21468/SciPostPhys.20.3.072},
	url={https://scipost.org/10.21468/SciPostPhys.20.3.072},
}

@article{PhysRevB.101.214302,
  title = {Validity of mean-field theory in a dissipative critical system: Liouvillian gap, $\mathbb{PT}$-symmetric antigap, and permutational symmetry in the $\mathit{XYZ}$ model},
  author = {Huybrechts, Dolf and Minganti, Fabrizio and Nori, Franco and Wouters, Michiel and Shammah, Nathan},
  journal = {Phys. Rev. B},
  volume = {101},
  issue = {21},
  pages = {214302},
  numpages = {21},
  year = {2020},
  month = {Jun},
  publisher = {American Physical Society},
  doi = {10.1103/PhysRevB.101.214302},
  url = {https://link.aps.org/doi/10.1103/PhysRevB.101.214302}
}

@article{PhysRevA.102.012219,
  title = {Nonequilibrium magnetic phases in spin lattices with gain and loss},
  author = {Huber, Julian and Kirton, Peter and Rabl, Peter},
  journal = {Phys. Rev. A},
  volume = {102},
  issue = {1},
  pages = {012219},
  numpages = {13},
  year = {2020},
  month = {Jul},
  publisher = {American Physical Society},
  doi = {10.1103/PhysRevA.102.012219},
  url = {https://link.aps.org/doi/10.1103/PhysRevA.102.012219}
}

@article{Temme_2013,
   title={Lower bounds to the spectral gap of Davies generators},
   volume={54},
   ISSN={1089-7658},
   url={http://dx.doi.org/10.1063/1.4850896},
   DOI={10.1063/1.4850896},
   number={12},
   journal={Journal of Mathematical Physics},
   publisher={AIP Publishing},
   author={Temme, Kristan},
   year={2013},
   month=dec }

@article{Kastoryano:2016feb,
    author = "Kastoryano, Michael J. and Brand{\~a}o, Fernando G. S. L.",
    title = "{Quantum Gibbs Samplers: The Commuting Case}",
    doi = "10.1007/s00220-016-2641-8",
    journal = "Commun. Math. Phys.",
    volume = "344",
    number = "3",
    pages = "915--957",
    year = "2016"
}

@article{Diehl2008,
  author  = {Diehl, S. and Micheli, A. and Kantian, A. and Kraus, B. and B{\"u}chler, H. P. and Zoller, P.},
  title   = {Quantum states and phases in driven open quantum systems with cold atoms},
  journal = {Nature Physics},
  volume  = {4},
  number  = {11},
  pages   = {878--883},
  year    = {2008},
  doi     = {10.1038/nphys1073}
}

@article{Diehl_2011,
   title={Topology by dissipation in atomic quantum wires},
   volume={7},
   ISSN={1745-2481},
   url={http://dx.doi.org/10.1038/nphys2106},
   DOI={10.1038/nphys2106},
   number={12},
   journal={Nature Physics},
   publisher={Springer Science and Business Media LLC},
   author={Diehl, Sebastian and Rico, Enrique and Baranov, Mikhail A. and Zoller, Peter},
   year={2011},
   month=oct, pages={971–977} }

@article{Bardyn_2013,
   title={Topology by dissipation},
   volume={15},
   ISSN={1367-2630},
   url={http://dx.doi.org/10.1088/1367-2630/15/8/085001},
   DOI={10.1088/1367-2630/15/8/085001},
   number={8},
   journal={New Journal of Physics},
   publisher={IOP Publishing},
   author={Bardyn, C-E and Baranov, M A and Kraus, C V and Rico, E and İmamoğlu, A and Zoller, P and Diehl, S},
   year={2013},
   month=aug, pages={085001} }

@article{Barreiro_2011,
   title={An open-system quantum simulator with trapped ions},
   volume={470},
   ISSN={1476-4687},
   url={http://dx.doi.org/10.1038/nature09801},
   DOI={10.1038/nature09801},
   number={7335},
   journal={Nature},
   publisher={Springer Science and Business Media LLC},
   author={Barreiro, Julio T. and Müller, Markus and Schindler, Philipp and Nigg, Daniel and Monz, Thomas and Chwalla, Michael and Hennrich, Markus and Roos, Christian F. and Zoller, Peter and Blatt, Rainer},
   year={2011},
   month=feb, pages={486–491} }

@article{Lin_2013,
   title={Dissipative production of a maximally entangled steady state of two quantum bits},
   volume={504},
   ISSN={1476-4687},
   url={http://dx.doi.org/10.1038/nature12801},
   DOI={10.1038/nature12801},
   number={7480},
   journal={Nature},
   publisher={Springer Science and Business Media LLC},
   author={Lin, Y. and Gaebler, J. P. and Reiter, F. and Tan, T. R. and Bowler, R. and Sørensen, A. S. and Leibfried, D. and Wineland, D. J.},
   year={2013},
   month=nov, pages={415–418} }

@article{Shankar:2013xif,
    author = "Shankar, S. and Hatridge, M. and Leghtas, Z. and Sliwa, K. M. and Narla, A. and Vool, U. and Girvin, S. M. and Frunzio, L. and Mirrahimi, M. and Devoret, M. H.",
    title = "{Autonomously stabilized entanglement between two superconducting quantum bits}",
    doi = "10.1038/nature12802",
    journal = "Nature",
    volume = "504",
    number = "7480",
    pages = "419--422",
    year = "2013"
}

@article{Leghtas_2015,
   title={Confining the state of light to a quantum manifold by engineered two-photon loss},
   volume={347},
   ISSN={1095-9203},
   url={http://dx.doi.org/10.1126/science.aaa2085},
   DOI={10.1126/science.aaa2085},
   number={6224},
   journal={Science},
   publisher={American Association for the Advancement of Science (AAAS)},
   author={Leghtas, Z. and Touzard, S. and Pop, I. M. and Kou, A. and Vlastakis, B. and Petrenko, A. and Sliwa, K. M. and Narla, A. and Shankar, S. and Hatridge, M. J. and Reagor, M. and Frunzio, L. and Schoelkopf, R. J. and Mirrahimi, M. and Devoret, M. H.},
   year={2015},
   month=feb, pages={853–857} }

@article{Kastoryano_2011,
   title={Dissipative Preparation of Entanglement in Optical Cavities},
   volume={106},
   ISSN={1079-7114},
   url={http://dx.doi.org/10.1103/PhysRevLett.106.090502},
   DOI={10.1103/physrevlett.106.090502},
   number={9},
   journal={Physical Review Letters},
   publisher={American Physical Society (APS)},
   author={Kastoryano, M. J. and Reiter, F. and Sørensen, A. S.},
   year={2011},
   month=feb }

@article{Plenio_1999,
   title={Cavity-loss-induced generation of entangled atoms},
   volume={59},
   ISSN={1094-1622},
   url={http://dx.doi.org/10.1103/PhysRevA.59.2468},
   DOI={10.1103/physreva.59.2468},
   number={3},
   journal={Physical Review A},
   publisher={American Physical Society (APS)},
   author={Plenio, M. B. and Huelga, S. F. and Beige, A. and Knight, P. L.},
   year={1999},
   month=mar, pages={2468–2475} }

@article{PRXQuantum.6.010361,
  title = {Quasiparticle Cooling Algorithms for Quantum Many-Body State Preparation},
  author = {Lloyd, Jerome and Michailidis, Alexios A. and Mi, Xiao and Smelyanskiy, Vadim and Abanin, Dmitry A.},
  journal = {PRX Quantum},
  volume = {6},
  issue = {1},
  pages = {010361},
  numpages = {30},
  year = {2025},
  month = {Mar},
  publisher = {American Physical Society},
  doi = {10.1103/PRXQuantum.6.010361},
  url = {https://link.aps.org/doi/10.1103/PRXQuantum.6.010361}
}

@article{bobkov2006modified,
  author  = {Bobkov, Sergey G. and Tetali, Prasad},
  title   = {Modified Logarithmic Sobolev Inequalities in Discrete Settings},
  journal = {Journal of Theoretical Probability},
  volume  = {19},
  number  = {2},
  pages   = {289--336},
  year    = {2006},
  doi     = {10.1007/s10959-006-0016-3}
}

@article{Zegarlinski1990LogSobolevIF,
  title={Log-Sobolev inequalities for infinite one dimensional lattice systems},
  author={Boguslaw Zegarlinski},
  journal={Communications in Mathematical Physics},
  year={1990},
  volume={133},
  pages={147-162},
  url={https://api.semanticscholar.org/CorpusID:123465656}
}

@article{PhysRevLett.125.230604,
  title = {Resolving a Discrepancy between Liouvillian Gap and Relaxation Time in Boundary-Dissipated Quantum Many-Body Systems},
  author = {Mori, Takashi and Shirai, Tatsuhiko},
  journal = {Phys. Rev. Lett.},
  volume = {125},
  issue = {23},
  pages = {230604},
  numpages = {6},
  year = {2020},
  month = {Dec},
  publisher = {American Physical Society},
  doi = {10.1103/PhysRevLett.125.230604},
  url = {https://link.aps.org/doi/10.1103/PhysRevLett.125.230604}
}

@article{PhysRevResearch.3.043137,
  title = {Metastability associated with many-body explosion of eigenmode expansion coefficients},
  author = {Mori, Takashi},
  journal = {Phys. Rev. Res.},
  volume = {3},
  issue = {4},
  pages = {043137},
  numpages = {7},
  year = {2021},
  month = {Nov},
  publisher = {American Physical Society},
  doi = {10.1103/PhysRevResearch.3.043137},
  url = {https://link.aps.org/doi/10.1103/PhysRevResearch.3.043137}
}

@misc{SM,
  note = {See the Supplemental Material for details on:
(A) a detailed analysis of the mixing time in the strong bulk-dissipation regime; 
(B) proofs of sufficient conditions under which the trace norm of the decaying eigenmode is bounded by $\mathrm{poly}(L)$ in both the strong- and weak-dissipation regimes as stated in Eq.\eqref{trace_norm_require_strong} and Eq.\eqref{trace_norm_require_weak};
(C) a clarification of two equivalent conventions for expressing the mixing time; 
(D) a simpler but stronger sufficient condition for fast and rapid mixing, based on bounding the trace-norm condition number of the full Liouvillian eigenbasis; 
(E) a discussion of why the number of decay eigenmodes $N_{\rm mode}$ is not expected to modify the rapid-mixing scaling in the generic situations considered in this work; and 
(F) a coarse-grained formulation of the mixing-time bound in terms of decay-rate bands;
and (G) how the Liouvillian skin effect produces an exponentially large spectral-projector prefactor and hence the scaling \(\tau_{\rm mix}\sim \Delta^{-1}(1+L/\xi)\).
}
}

@article{Zhou_2024_LSM,
   title={Reviving the Lieb–Schultz–Mattis theorem in open quantum systems},
   volume={12},
   ISSN={2053-714X},
   url={http://dx.doi.org/10.1093/nsr/nwae287},
   DOI={10.1093/nsr/nwae287},
   number={1},
   journal={National Science Review},
   publisher={Oxford University Press (OUP)},
   author={Zhou, Yi-Neng and Li, Xingyu and Zhai, Hui and Li, Chengshu and Gu, Yingfei},
   year={2024},
   month=aug }

@Article{10.21468/SciPostPhysCore.6.1.023,
	title={{Generalized Lindblad master equation for measurement-induced phase transition}},
	author={Yi-Neng Zhou},
	journal={SciPost Phys. Core},
	volume={6},
	pages={023},
	year={2023},
	publisher={SciPost},
	doi={10.21468/SciPostPhysCore.6.1.023},
	url={https://scipost.org/10.21468/SciPostPhysCore.6.1.023},
}

@article{Brand_o_2015,
   title={Area law for fixed points of rapidly mixing dissipative quantum systems},
   volume={56},
   ISSN={1089-7658},
   url={http://dx.doi.org/10.1063/1.4932612},
   DOI={10.1063/1.4932612},
   number={10},
   journal={Journal of Mathematical Physics},
   publisher={AIP Publishing},
   author={Brandão, Fernando G. S. L. and Cubitt, Toby S. and Lucia, Angelo and Michalakis, Spyridon and Perez-Garcia, David},
   year={2015},
   month=Oct }

@article{Temme_2010,
   title={The  $\chi^2$-divergence and mixing times of quantum Markov processes},
   volume={51},
   ISSN={1089-7658},
   url={http://dx.doi.org/10.1063/1.3511335},
   DOI={10.1063/1.3511335},
   number={12},
   journal={Journal of Mathematical Physics},
   publisher={AIP Publishing},
   author={Temme, K. and Kastoryano, M. J. and Ruskai, M. B. and Wolf, M. M. and Verstraete, F.},
   year={2010},
   month=Dec }

@article{PhysRevResearch.4.L022039,
  title = {Space-time duality between quantum chaos and non-Hermitian boundary effect},
  author = {Zhou, Tian-Gang and Zhou, Yi-Neng and Zhang, Pengfei and Zhai, Hui},
  journal = {Phys. Rev. Res.},
  volume = {4},
  issue = {2},
  pages = {L022039},
  numpages = {6},
  year = {2022},
  month = {May},
  publisher = {American Physical Society},
  doi = {10.1103/PhysRevResearch.4.L022039},
  url = {https://link.aps.org/doi/10.1103/PhysRevResearch.4.L022039}
}
\end{document}